\documentclass[10pt]{article}
\usepackage{amsmath,amsfonts,amssymb,amsthm,bbm}
\usepackage{pdfsync}
\usepackage{graphicx}
\usepackage[latin1]{inputenc}
\usepackage{hyperref}
\usepackage[all]{xy}
\usepackage{stmaryrd}
\usepackage[percent]{overpic}
\usepackage{dsfont}
\usepackage[normalem]{ulem}

\newcommand{\bit}{\begin{itemize}}
\newcommand{\eit}{\end{itemize}}
\newcommand{\bd}{\begin{description}}
\newcommand{\ed}{\end{description}}

\newcommand{\bc}{\begin{center}}
\newcommand{\ec}{\end{center}}

\newcommand{\C}{{\mathbb C}}

\newcommand{\R}{{\mathbb R}}
\newcommand{\Z}{{\mathbb Z}}

\newcommand{\SU}{\mathrm{SU}}

\newcommand{\SL}{\mathrm{SL}}
\newcommand{\SO}{\mathrm{SO}}

\renewcommand{\sl}{{\mathfrak{sl}}}

\newcommand{\be}{\begin{equation}}
\newcommand{\ee}{\end{equation}}
\newcommand{\bea}{\begin{eqnarray}}
\newcommand{\eea}{\end{eqnarray}}
\newcommand{\bs}{\begin{subequations}}
\newcommand{\es}{\end{subequations}}
\newcommand{\nn}{\nonumber}

\newcommand{\w}{\wedge}
\newcommand{\sgn}{\mathrm{sgn}}

\newcommand{\f}{\frac}
\newcommand{\tl}{\tilde}
\def\p{\partial}
\newcommand{\Id}{\mathbbm{1}}

\newcommand{\re}{\mathrm{Re}}
\newcommand{\im}{\mathrm{Im}}

\newcommand{\scr}{\scriptscriptstyle\rm}

\newcommand{\ra}{\rangle}
\newcommand{\la}{\langle}
\newcommand{\bra}[1]{\langle {#1}|}
\newcommand{\ket}[1]{|{#1}\rangle}
\newcommand{\sbra}[1]{[ {#1}|}
\newcommand{\sket}[1]{|{#1} ]}

\newcommand{\vect} [2] {\left ( \begin{array}{c} #1 \\ #2\end{array} \right ) }
\newcommand{\mat} [4] {\left ( \begin{array}{cc} #1 & #2 \\ #3 & #4\end{array} \right ) }

\renewcommand{\a}{\alpha} \renewcommand{\b}{\beta} \newcommand{\g}{\gamma}
\renewcommand{\d}{\delta}  \newcommand{\eps}{\epsilon}  \newcommand{\z}{\zeta}
 \renewcommand{\th}{\theta}      \renewcommand{\l}{\lambda}
\let\m=\mu    \renewcommand{\r}{\rho} \newcommand{\s}{\sigma}       \let\om=\omega
\let\G=\Gamma \let\D=\Delta   \let\L=\Lambda  \let\Om=\Omega

\newcommand{\norm}[1]{|\!| #1 |\!|}  

\usepackage{xcolor} 

\usepackage[autostyle, english = british]{csquotes}
\MakeOuterQuote{"}

\newcommand{\Kp}{K'}
\newcommand{\magg}{boosted orientation }
\newcommand{\cc}{{\scr (c)}}
\newcommand{\Pcc}{{\scr (\bar{c})}}
\newcommand{\rr}{r}

\begin{document}

\title{\bf Asymptotics of $\SL(2,\C)$ coherent invariant tensors}

\author{\Large{Pietro Don\`a$^a$, Marco Fanizza$^b$, Pierre Martin-Dussaud$^{a,c}$ and Simone Speziale$^a$}
\smallskip \\ 
\small{$^a$ Aix Marseille Univ., Univ. de Toulon, CNRS, CPT, UMR 7332, 13288 Marseille, France} \\
\small{$^b$NEST, Scuola Normale Superiore and Istituto Nanoscienze-CNR, I-56126 Pisa, Italy}\\
\small{$^c$ IGC and Department of Physics, PennState University, University Park, Pennsylvania 16802, USA} }
\date{\today}

\maketitle

\begin{abstract}
\noindent We study the semiclassical limit of a class of invariant tensors for infinite-dimensional unitary representations of SL(2,C) of the principal series, corresponding to generalized Clebsch-Gordan coefficients with $n\geq3$ legs.
We find critical configurations of the quantum labels with a power-law decay of the invariants. They describe 3d polygons that can be deformed into one another via a Lorentz transformation. This is defined viewing the edge vectors of the polygons are the electric part of bivectors satisfying a (frame-dependent) relation between their electric and magnetic parts known as $\g$-simplicity in the loop quantum gravity literature. The frame depends on the SU(2) spin labelling the basis elements of the invariants.
We compute a saddle point approximation using the critical points and provide a leading-order approximation of the invariants. The power-law is universal if the SU(2) spins have their lowest value, and $n$-dependent otherwise.
As a side result, we provide a compact formula for $\g$-simplicity in arbitrary frames. The results have applications to the current EPRL model, but also to future research aiming at going beyond the use of fixed time gauge in spin foam models.
\end{abstract}

\tableofcontents

\section{Introduction}
%

The recoupling theory of SU(2) contains beautiful formulas relating the semiclassical limit of Clebsch-Gordan coefficients and invariant tensors
to Euclidean geometry. The derivation of these formulas started with the seminal work of Wigner followed up by Ponzano and Regge (see e.g. \cite{wigner1959group,PonzanoRegge,Schulten:1971yv}). Recent developments include higher order corrections (e.g. \cite{Bonzom:2008xd, Dupuis:2009sz,Han:2020fil}), extensions to invariant tensors associated to more complicated graphs (e.g. \cite{Haggard:2009kv,BarrettSU2,IoSU2asympt}), and to other groups (e.g. \cite{Davids:1998bp,taylor20066j,Barrett:2002ur,Freidel:2002mj,Krasnov:2005fu}). Following this line of research, we study in this paper the semiclassical limit of $\SL(2,\C)$ Clebsch-Gordan coefficients for the infinite dimensional unitary representations of the principal series.
Our motivation comes from models of quantum gravity, where $\SL(2,\C)$ appears as the gauge group related to the Lorentz symmetry of a local inertial frame. However, our results are of a general mathematical nature, and can be applied to any system with an $\SL(2,\C)$ symmetry. 

The most common $\SL(2,\C)$ invariant studied in quantum gravity models is the equivalent of the SU(2) $15j$ symbol, which provides the 4-simplex spin foam vertex amplitude for loop quantum gravity (LQG). It is defined by four group averagings of ten infinite-dimensional unitary representation matrices of the principal series. The matrices are written in Naimark's basis, which diagonalizes the canonical SU(2) subgroup. Various definitions of the amplitude are available in the literature. The former Barrett-Crane (BC) model included only one family of irreducible representations (irreps) \cite{BarrettCraneLor}, and its asymptotics was studied in \cite{Barrett:2002ur,Freidel:2002mj}. The more recent Engle-Pereira-Rovelli-Livine (EPRL) model \cite{EPRL,FK}, currently the state of the art for the spin foam formalism, uses both families but the SU(2) spin is fixed at the lowest value. Its asymptotics was studied in \cite{BarrettLorAsymp}, and led to many applications (e.g. \cite{Riello13,HanZhangLor,Haggard:2015yda,Engle:2015zqa,Kaminski:2017eew,Liu:2018gfc,Dona:2020tvv}). A key technical tool for the geometric interpretation is the use of SU(2) coherent states and coherent intertwiners \cite{LS}.
Just like the semiclassical limit of SU(2) invariants is described by Euclidean geometry, the semiclassical limit of $\SL(2,\C)$ invariants is described by Minkowskian geometry.
In both BC and EPRL models, the asymptotics of the 4-simplex amplitude are related to the Minkowskian geometry of Lorentzian 4-simplices and the Lorentzian Regge action. The Kaminski-Kisielowski-Lewandowski (KKL) model provides an extension of EPRL to arbitrary vertex amplitudes in \cite{KKL}. Its asymptotics are related to Lorentzian 4d polytopes, as well as to more general objects describing conformal twisted geometries \cite{BahrSteinhaus15,Bahr:2018vvq,Dona:2020yao}. A generalized notion of Regge action emerges in all these cases.

The aim of this paper is to go beyond the case of lowest SU(2) spin used in the EPRL amplitudes. To that end, we remove the additional complexity associated with a graph structure, and we consider the simplest invariant tensors: a single group averaging of $n$ infinite-dimensional unitary representation matrices of the principal series.
These invariants correspond to products of two (generalized) Clebsch-Gordan coefficients, namely the analogue of (generalized) Wigner's $(3jm)$ symbols for SU(2). 
These asymptotics were not considered before (with the exception of the lowest SU(2) spins and no coherent states \cite{Puchta:2013lza}). Our work fills this gap.
A specific motivation to study these tensors comes from the technique used to evaluate arbitrary $\SL(2,\C)$ invariants introduced in \cite{Boosting}. 
Applied to spin foam models, this method leads to a decoupling between vertex amplitudes which are purely SU(2), and edge amplitudes which are one-dimensional integrals over a boost parameter. It can be applied to analytic and numerical studies \cite{noiGen,Dona:2018pxq,Dona:2018nev,noiLor}, 
and stimulates further analysis of the boost integrals appearing in the edge amplitudes, now commonly referred to as booster functions. 
The invariants studied in this paper are booster functions in a basis of SU(2) coherent states.
The use of coherent states gives a finer structure to the labels of the invariants, and allows us to classify them in geometric terms. In particular, the SU(2) spins and the coherent states define norms and directions of vectors in $\R^3$, and the semiclassical limit of the invariants endows them with an interpretation in Minkowski space. 

In this paper we study the asymptotic behaviour of the coherent invariant tensors, or coherent booster functions, with all quantum numbers homogeneously large.
We show that the generic invariant has an exponential fall-off unless the quantum labels and coherent states satisfy measure-zero constraints. In this case, there are critical points for the integration variables that lead to a power-law fall-off.
There are two types of constraints that the critical data must satisfy. First, each set of data corresponding to the rows and columns of the invariants must describe a polygon in 3d, or equivalently for non-coplanar configurations, a 3d convex polyhedron. Second, the two sets must be related by a certain Lorentz transformation. This is defined through a map from the coherent states to bivectors. This map emerges from the critical point equations, but can be anticipated looking at the bivector built from the expectation values of the Lorentz generators. These satisfy a proportionality between boosts and rotations (or equivalently between magnetic and electric parts) known as $\g$-simplicity.

The notion of $\g$-simplicity is familiar from the case of lowest SU(2) spins from the EPRL model.
In this case, the expectation values of the Lorentz generators define bivectors which are $\g$-simple in the frame of the canonical time direction. This specific direction is built in the problem from the use of Naimark's basis.
The main novelty revealed by our analysis is that also for arbitrary spins it is possible to identify $\g$-simple bivectors, this time in a boosted frame. The boosted frame depends on the SU(2) spins, as well as on the flags of the spinors describing the boundary coherent states. 
This leads to a particular intricacy of the critical point equations, which are more complicated to solve than in the case of lowest spins.
The critical behaviour requires a compatibility condition between the frames of $\g$-simplicity of the two sets of boundary data. Otherwise, there is no Lorentz transformation possible between the two, and thus there are no critical points. This compatibility condition leads to complicated 
equations. When one set of labels has all spins at lowest value, we were able to solve the equations explicitly. 
This is actually the case directly relevant for applications to the EPRL model. In the general case, the equations can be solved numerically, but we don't have an explicit algebraic solution. Nonetheless, we have clarified their meaning and solutions in purely geometric terms. These results 
thus provide an elegant geometric interpretation of the coherent invariant tensors. 

For the critical configurations identified, the power-law fall-off of the invariants is universal for the case of lowest spins, and $n$-dependent otherwise. Furthermore, we find a unique critical point for lowest spins, and two distinct critical points when only the spins in one set are lowest. 
We complete our analytical work with numerical tests, and provide explicit examples of critical data and plots confirming our leading order asymptotic formula. The numerical evaluations are done using the $C$ code \cite{Dona:2018nev} and its newer version \cite{Gozzini-to-appear?}. Supplementary Mathematica notebooks can be found in \cite{repo}.

We use Minkowski signature $-+++$ with indices $I=0,\ldots 3$, and $i=1,2,3$ for space indices. The antisymmetric symbol is taken with conventions $\eps_{0123}=1$, and the Hodge dual is defined as 
$(\star \om^{(p)})_{\m_1..\m_{n-p}} := (1/{p!}) \om^{(p)}{}^{\a_1..\a_p} \eps_{\a_1..\a_p\m_{1}..\m_{n-p}}$.

\section{Preliminaries}\label{SecPreliminaries}

In this Section, we summarize useful background material. In particular, we describe the factorization property of coherent states, their map to $\g$-simple bivectors, and their Lorentz transformations.

\subsection{$\SL(2,\C)$ unitary irreps of the principal series}

We denote $\hat J^{IJ}$ the  $\SL(2,\C)$ generators, with ${\hat L}^i=- \frac{1}{2} \eps^{ijk}\hat J_{jk}$ the rotations and ${\hat K}^i=\hat J^{0i}$ the boosts, and algebra
\be
[{\hat L}^i,\hat  {L}^j]=i\eps^{i j}_{\ \ k}\hat L^k , \qquad [{\hat K}^i,\hat  { K}^j]=-i\eps^{i j}_{\ \ k}\hat J^k, \qquad [{\hat L}^i,\hat  { K}^j]=i\eps^{i j}_{\ \ k}\hat K^k.
\ee
Unitary irreducible representations (irreps) of the principal series are labelled by a pair $(\r \in \R,k\in \Z/2)$ \cite{GelfandLorentz, Naimark,Ruhl}, with  Casimirs taking values\footnote{Our conventions follow those of the monograph by Ruhl \cite{Ruhl}, with $\r=\r_{\rm Ruhl}/2, k=-m_{\rm Ruhl}/2$. } 
\be
{\cal C}_{(1)}:=\vec {\hat L}^2-\vec {\hat K}^2 =( \r^2 - k^2 -1 ), \qquad {\cal C}_{(2)}:=\vec {\hat K}\cdot \vec {\hat L}= 2 \rho k. 
\ee
These irreps can be realized on a space of (non-holomorphic) homogenous functions of two complex variables,
\be
F^{(\r,k)}(\l z^A) = \lambda^{k-1+i \rho}\bar{\lambda}^{-k-1+i \rho}\, F^{(\r,k)}(z^A), \qquad \l\in\C, \qquad A=0,1.
\ee
From this formula we can easily see that $(\rho,k)$ and $(-\rho,-k)$ are related by complex conjugation. The group action is given by   
\be
h\triangleright F(z^A) = F(h^{\scr T} z^A), \qquad h\in\SL(2,\C),
\ee
where $h^{\scr T}$ is the transpose of $h$, and $h z^A:=h^A{}_B z^B$ is a shorthand notation for matrix multiplication. It will often be convenient to think of $z^A\in\C^2$ as a spinor.
Spinorial indices can be raised and lowered with the antisymmetric tensor $\eps_{AB}$, and we fix conventions $z_A:=z^B\eps_{BA}$, $\eps_{01}=\eps^{01}=1$.

The scalar product is defined choosing a path $\G:\C P^1\mapsto \C^2$ crossing once and once only each complex line through the origin of $\C^2$, \be
(F,F'):=\int_{\G} d\m(z^A) \overline{ F(z^A) }F'(z^A), \qquad d\m(z^A):=\f i2 z_A d z^A\w \bar z_{\dot A} d \bar z^{\dot A}.
\ee
The scaling of the Lorentz-invariant measure $d\m(z^A)$ guarantees that the whole integrand is homogeneous of degree zero, and thus the integration is independent of the choice of $\G$.
Geometrically, the choice of path can be seen as a choice of section for the tautological bundle $\C^2\simeq \C P^1\times \C$.

We will use Naimark's orthonormal basis $\ket{\r,k;j,m}$, which diagonalizes the operators $\hat L^2$ and $\hat L_z$ of the matrix subgroup SU(2), and is given explicitly by
\begin{align}\label{Naimark}
&F^{(\r,k)}_{jm}(z^A)\equiv \la z^A\ket{\r,k;j,m} 
:= \sqrt{\frac {d_j}{\pi}} \f{1}{\norm{z}^{2(1-i\rho)}}D^{(j)}_{m,-k}\big(g(z^A)\big),
\qquad  j\geq |k|, \quad -j\leq m \leq j,
\end{align}
where $\norm z^2:=\d_{A\dot A}z^A z^{\dot A}$ is the SU(2)-invariant Hermitian norm and $D^{(j)}(g)$ the SU(2) Wigner matrices, and
\be
g(z^A):=\f1{\norm z}\mat{\bar z^1}{z^0}{-\bar z^0}{z^1} \in \SU(2).
\ee

The matrix elements of the infinite-dimensional representation matrices associated to Naimark's orthonormal basis can be constructed from the scalar products,
\be\label{Dh}
D^{(\r,k)}_{jmln}(h) := \int_{\C P^1} d\m(z^A)  \overline{F^{(\r,k)}_{jm}(z^A)} \, F^{(\r,k)}_{ln}(h^{\scriptscriptstyle\rm T}z^A),
\ee
and the unitarity of the representation follows from the Lorentz-invariance of the integration measure. 
In the following, we will always assume $k > 0$.
For more details on these irreps and their use in LQG, see \cite{Martin-Dussaud:2019ypf}.

\subsection{SU(2) Coherent states}

The label $m$ of the orthonormal basis can be given a geometric interpretation as projection along the $\hat z$ axis of the rotation generator $\hat L^i$. As such, each state has minimal information about the direction of the rotation generator. A sharper characterization can be obtained using SU(2) coherent states, which replace the discrete label $m$ with two continuous parameters marking a point on the sphere. The rotation generator is then peaked on the direction identified by this point. To minimize the uncertainty, the SU(2) coherent states are defined as rotations of 
minimal weight state,
 \be\label{SU2cs}
\ket{j,\z}:=D^{(j)}\big( n(\z)\big)\ket{j,-j}, 
\ee
where 
\be\label{Hs}
n(\z) := \f1{\sqrt{1+|\z|^2}}\left( \begin{array}{cc} 
1 & \z \\  -\bar\z & 1
\end{array}\right)
\ee
is the Hopf section for the $\SU(2) \simeq S^2\times S^1$ fibration. Writing $\z=-\tan(\th/2) e^{-i\phi}$ as a stereographic projection from the south pole, 
in the fundamental representation $j=1/2$ we have
\be\label{plat}
\ket{\z}:=\ket{\tfrac12,\z}=\vect{-\sin\f\th2 e^{-i\phi}}{\cos\f\th2}.
\ee
We can think of it as spinor $\z^A\in\C^2$ with unit norm and $\arg \z^1=0$. The map from \eqref{plat} to a unit vector in $\R^3$ is provided by
\be\label{PauliEV}
\bra{\z}\vec\s\ket{\z} = -\vec n:=-(\sin\th\cos\phi,\sin\th\sin\phi,\cos\th),
\ee
where $\vec\s$ are the Pauli matrices. In a generic irrep $j$, 
$\bra{j,\z}\vec{\hat L}\ket{j,\z} = -j\vec n $. This family of states minimizes the uncertainty on the direction of the generator.

The SU(2) coherent states can be embedded in the unitary irreps of $\SL(2,\C)$ as follows,
 \be\label{SL2Ccs}
\ket{\r,k;j,\z}:=D^{(j)}\big( n(\z)\big)\ket{\r,k;j,-j}.
\ee
They have some semiclassical properties:
the expectation values (EVs) of rotations and boosts point in the direction identified by the label of the state, 
\be\label{LKev}
\vec L:= \bra{\r,k;j,\z} \vec {\hat L} \ket{ \r,k;j,\z} = -j\vec n, \qquad \vec K:= \bra{\r,k;j,\z} \vec{\hat{K}} \ket{ \r,k;j,\z} = -\f{\r k}{j+1} \vec n,
\ee
with minimal uncertainty in the direction of the rotation generators -- but not of the boost generators.
\footnote{This $\SL(2,\C)$ embedding of SU(2) coherent states should not be confused with $\SL(2,\C)$ coherent states in the sense of \cite{Perelomov}. The latter are only defined for the $k=0$ irreps, as the 3-parameter family $D^{(\r,k)}(b)\ket{\r,0;0,0}$, where $b$ is a pure boost.} 

We can now define the overcomplete basis of SU(2) coherent states for the homogeneous realization, given by
\begin{align}\label{Fcs}
F^{(\r,k)}_{j\z}(z^A) & := \sum_m \la j,m \ket{j,\z} F^{(\r,k)}_{jm}(z^A)
= \sqrt{\frac{d_j}{\pi}}\, \norm{z}^{2(i\r-1-j)} \left(\begin{array}{c}2j \\j+k \end{array} \right)^{1/2} [\z\ket{\bar z}^{j-k} \bra{\bar z}\z\ra^{j+k}.
\end{align}
In the equation above we introduced the following Dirac-like short-hand notation for spinors and their Hermitian and Lorentz-invariant bi-linears:
\begin{align}
& \ket{z}=z^A, \qquad \bra{z}=\d_{A\dot A}\bar z^{\dot A}, \qquad [z|=z_A, \qquad |z]=\d^{A\dot A}\bar z_{\dot A}, \\\label{due}
& \bra{z}w\ra=\d_{A\dot A}\bar z^{\dot A}w^A =
[w|z], \qquad [z\ket{w}=\eps_{AB}z^A w^B. 
\end{align}
The second equality in \eqref{Fcs} is most easily proved removing the resolution of the identity over $m$ and multiplying the matrices $g(z^A)$ and $n(\z)$ together. It re-expresses the sum over magnetic indices in terms of scalar products in the fundamental representation. For the lowest spin $j=k$ there is a single scalar product, and this is the situation occurring in  \cite{BarrettLorAsymp}. Equation \eqref{Fcs} shows that the factorization holds for any spins, with the two different scalar products appearing. This factorization property of coherent states plays a crucial role in setting up the saddle point analysis of the invariant tensors. 

Since we are fixing once and for all the global phase of spinor via \eqref{plat}, we can identify the label of the coherent state $\zeta$ directly with the unit norm vector $\vec n$. It is also possible to relax this condition, and allow for an arbitrary unit-norm spinor $z^A$ as label. This  provides a redundant overcomplete basis
of coherent states.\footnote{There is an analogue situation in canonical LQG, where one can switch from the usual heat-kernel coherent states -- or the twisted geometry coherent states -- to a redundant overcomplete basis parametrized by  spinors, see the comparison made in \cite{Calcinari:2020bft}.} Keeping in mind this possibility can be useful to understand some structures of the amplitudes.

Thanks to the factorization \eqref{Fcs}, the group matrix elements in the coherent state admit a compact expression,
\begin{align}\label{Dcs}
D^{(\r,k)}_{j\tl\z l {\z}}(h) &:= 
\int_{\C P^1} d\m(z^A)  \overline{F^{(\r,k)}_{j\tl\z}(z^A)} \, F^{(\r,k)}_{l\z}(h^{\scriptscriptstyle\rm T}z^A)
= {\cal N}^k_{jl} \int_{\C P^1} \f {d\m(z^A)}{\norm{\bar z}^{2} \norm{h^{\dagger} \bar z}^{2} } e^{\tl s(h,z;\z,\tl\z)},
\end{align}  
with numerical prefactor 
\be\label{defcN}
{\cal N}^k_{jl}:=\frac{\sqrt{d_j d_l}}{\pi} \binom{2j}{j+k}^{1/2} \binom{2l}{l+k}^{1/2},
\ee
and complex `action'
\begin{align}\label{defs}
\tl s(h,z;\z,\tl\z)&:=\log\f{ \bra{\bar z}{\tl \z}]^{j-k} \bra{{\tl\z}}\bar z\ra^{j+k} }{\norm{\bar z}^{2(j+i\r)}} 
+ \log\f{ [\z \ket{ h^{\dagger}\bar z}^{l-k} \bra{ h^{\dagger} \bar z } \z \ra^{l+k}}{\norm{h^{\dagger}\bar z}^{2(l-i\r)}}.
\end{align}
In defining \eqref{defs} we have chosen to put the group element $h$ only in the scalar products involving the $\z$ spinor. Given the Lorentz invariance of the measure $\m(z^A)$, $h$ can be equivalently placed in the scalar products with the $\tl\z$ spinors, as $(h^{\dagger})^{-1}$. Or if we have $h=h_1^{-1}h_2$ like in more complicated graph invariants, we can write $h_1$ and $h_2$ respectively in each set of scalar products.

\subsection{$\g$-simplicity and Lorentz transformations}
\label{sec:gsimple}
The expectation values \eqref{LKev} provide a 2-to-1 map from the coherent states \eqref{SL2Ccs} to a special class of bivectors, which plays an important role in the semiclassical limit of the invariant tensors. The map depends on the spin weight $j$. Consider first the lowest value, $j=k$. In this case, the expectation values \eqref{LKev}  are related by 
${\vec K - (\r/k) \vec L} =O(k^{-1})$. In the following, we will be interested in the limit of large labels, and discard all factors $O(k^{-1})$.\footnote{If one wants to keep them, it is possible to work with the parameter $\g':=\r/(k+1)$, and move the measure terms inside the definition of the action, so to have $\g'$ appear in the critical point equations.} In the large spin limit,
\be\label{gsimp}
\vec K = \g \vec L, \qquad \g:=\f \r k.
\ee
This property will be referred to as \emph{$\g$-simplicity}, and it is the cornerstone of the EPRL model \cite{EPRL}. 
As a vectorial equation, it has one Lorentz-invariant component, and two components depending explicitly on the time-like canonical direction $t^I=(1,0,0,0)$.
If we define the bivector $J^{IJ}$ with $J^{0i}=K^i$ and $J^{ij}=-\eps^{ij}{}_{k}L^k$, then $B^{IJ}:= (\Id-\g\star)J^{IJ}$ is simple, since $B^{0i}=0$ and thus $\eps_{IJKL}B^{IJ}B^{KL}=0$. This is the Lorentz-invariant content of \eqref{gsimp}. By $\g$-simplicity, we will mean the set of all three (frame-dependent) conditions.
In matrix form,
\be
J^{IJ} = k \begin{pmatrix}
0 & - \g n_x & - \g n_y & - \g n_z \\
\g n_x & 0 & n_z &  - n_y \\ 
\g n_y & - n_z & 0 & n_x \\ 
\g n_z & n_y & - n_x& 0
\end{pmatrix},
\qquad
B^{IJ} = (1+\g^2) k \begin{pmatrix}
0 & 0 & 0 & 0 \\ 
0 & 0 & n_z &  - n_y \\
0  & -n_z & 0 & n_x \\
0 & n_y & - n_x & 0
\end{pmatrix}.
\ee
Therefore, we can think of $-k\vec n$ as the electric part of a `$\g$-simple' bivector $J^{IJ}$. We also remark that the self-dual part of such a bivector generates a four-screw (a boost accompanied by a rotation in the same direction) in the direction of $\vec n$, since
\be\label{Pisimple}
\vec \Pi:=\vec L+i \vec K = -(k+i\r )\vec n =  (k+i\r )\bra{\z} \vec{\s} \ket{\z}. 
\ee
If $h\in\SL(2,\C)$, we have $\L=T^{(1/2,1/2)}(h)$ (we use $T$ instead of $D$ to distinguish the finite-dimensional irreps from the infinite-dimensional ones), and 
\begin{align}\label{defT10} 
 \vec{\Pi}_\L:=& T^{(1,0)}(h) \vec{\Pi} = (k+i\r )\bra{\z}h^{-1}\vec{\s} h\ket{\z}.
\end{align}
This map provides a 2-to-1 homomorphism between (proper orthochronus) Lorentz transformations $\L$ on the four-screw generators and $\SL(2,\C)$ transformations on the spinors.\footnote{The reader may be familiar with the map from spinors to the 4d space of null and simple bivectors. This is not the map used here. The space of four-screw bivectors is also 4d, but it is being spanned by the ratio of the spinor's components plus the two irrep parameters $\r$ and $k$. In the first case, both Casimirs are vanishing. In the second case, the value of the Casimirs is arbitrary.}
The explicit form of the transformation \eqref{defT10} can be conveniently written using the polar decomposition 
\begin{subequations}
\begin{align}
& h=\rr b, && T^{(1,0)}(h)=RB, \\
& \rr = e^{i \th \vec{v}\cdot \frac{\vec{\s}}{2}},&& R = \cos\th \,\Id - \sin\th \, \star \vec v   + (1-\cos\th) \vec v \otimes \vec v, \label{rodrotation} \\
& b = e^{\eta \vec{u}\cdot \frac{\vec{\s}}{2}},  && B = \cosh \eta \,\Id + i \sinh \eta \, \star \vec u   + (1-\cosh \eta) \vec u \otimes \vec u. \label{rodboost}
\end{align}
\end{subequations}
Taking real and imaginary parts of \eqref{defT10} and using \eqref{gsimp}, we find
\begin{align}
& \vec L\mapsto \vec L_\L=T_\g(h) \vec L, \qquad
 \vec K\mapsto \vec K_\L=T_{-1/\g}(h)\vec K, \\\label{Tgdef}
 & T_\g(h) = R \big( \re(B) -\g \im(B) \big)
 = R\left(\cosh \eta \,\Id - \g\sinh \eta \, \star \vec u   + (1-\cosh \eta) \vec u \otimes \vec u  \right).
\end{align}
We remark that $T_\g(h)$ is \emph{not} a representation of $\SL(2,\C)$,
and it shouldn't since the electric-magnetic decomposition is not invariant.
We can decompose it as rotations times a dilation,
\begin{equation}
\label{decomposition}
T_\g(h) = R\, R_{\vec u}(\psi) \, D_\gamma.
\end{equation}
Here $R_{\vec u}(\psi)$ is a rotation in the direction $\vec{u}$ of an angle 
\be
\label{psiangle}
\psi(\g,\eta) = \arccos\big( d^{-1}_\g(\eta) \cosh(\eta) \big), \qquad d_\g(\eta):=\sqrt{\cosh^2\eta+\g^2\sinh^2\eta}, 
\ee
and $D_\gamma$ is a dilation in the plane orthogonal to $\vec{u}$, 
\begin{equation}
\label{dilation}
 D_\gamma = d_\g(\eta) \,\Id + \left(1-d_\g(\eta)\right) \, \vec u \otimes \vec u \ .
\end{equation}
In the particular case $\g=0$ the simple bivector has vanishing magnetic part and $\psi(0,\eta)=0$, but $T_0(h)$ is still not a representation.

The property of $\g$-simplicity is covariant under Lorentz transformations.
This can be seen explicitly writing \eqref{gsimp} in covariant terms as
\be\label{simpcov}
t_I (\Id-\g\star)J^{IJ} = 0, \qquad t^I=(1,0,0,0).
\ee
Then $J_\L^{IJ}:= \L^I{}_K J^{KL} \L^J{}_L$ is again $\g$-simple, but with respect to a new time-like direction $N^I:=\L^I{}_J t^J$.
According to \eqref{simpcov}, its canonical generators satisfy
\begin{align}
\cosh \eta (\vec K-\g \vec L) = \sinh \eta \, (R^{-1}\vec u) \times (\vec L+\g \vec K).
\end{align}
Or in terms of the boosted electric and magnetic parts,
\be\label{Boostedgsimp}
\vec K_\L = \g T_{-1/\g}(h)T_{\g}(h)^{-1} \vec L_\L.
\ee
This is the $\g$-simplicity condition \eqref{gsimp} in an arbitrary frame.

Consider now the case with $j>k$. The expectation values $(\vec L,\vec K)$ do not satisfy the condition of $\g$-simplicity, because their proportionality is $j$-dependent. However, it is always possible to construct a  $\g$-simple bivector keeping $\vec L$, and looking for a new $\vec \Kp$ that would preserve the
(large-spin limit)  Casimir identities 
\begin{equation}\label{Casimirs}
\vec{\Kp}{}^2-\vec{L}^2=\r^2-k^2, \qquad \vec \Kp\cdot \vec L = \r k, \qquad \vec{L}^2=j^2.
\end{equation}
To that end, recall that a unit-norm spinor defines an orthonormal basis of $\R^3$ given by $(\vec n, \vec F, \vec n\times \vec F)$, where
\begin{equation}
\label{evsigma}
\bra{\z}\vec \s\ket{\z} = -\vec n, \qquad [\z|\vec \s\ket{\z} = i \vec F+\vec n\times \vec F.
\end{equation}
The `flag vector' $\vec F$ depends explicitly on the spinor's flag, namely that global phase information that is lost in the projection to $\vec n\in S^2$
(recall Penrose's picture of a spinor as a null pole plus null flag).
Using this basis, the most general bivector $(\vec L,\vec\Kp)$ satisfying \eqref{Casimirs} is given by
\begin{equation}
\label{KFF}
\vec \Kp = -\f{\r k}{j}\vec n + \frac{\sqrt{j^2-k^2}}{j} (-j\vec{p} + \r \vec{q}) = \frac{j+1}{j} \vec K + \frac{\sqrt{j^2-k^2}}{j} (-j\vec{p} + \r \vec{q}),
\end{equation}
where $(\vec{p},\vec{q})$ are obtained by an SO(2) rotation of $(\vec F, \vec n\times\vec F)$ with arbitrary parameter.
The new bivector can be written as a Lorentz transformation of (the generator of) a four-screw, 
\be\label{Piprime}
\vec{\Pi}' = \vec L + i \vec \Kp = -T^{(1,0)}(g) (k +i \r)\vec{n} =  (k+i\r )\bra{\z}g^{-1}\vec{\s} g\ket{\z},
\ee
with $g$ a unique four-screw up to a further four-screw in the $\vec n$ direction: 
\begin{align}
& g = \exp \left(\om \frac{\r\vec p + j \vec q}{\sqrt{j^2+\r^2}}\cdot\f{\vec \s}2\right) \, \exp \left((x+iy) \vec n\cdot\f{\vec \s}2\right), \label{4screw}
\\
& \omega = \sinh^{-1}\left(\frac{1}{j}\frac{\sqrt{\left(j^{2}-k^{2}\right)\left(j^{2}+\rho^{2}\right)}}{k+i\rho}\right) =
 \cosh^{-1}\f{j^2+i k\r}{j(k+i\r)} \in \C,
\end{align}
and $(x,y)$ arbitrary. Comparing \eqref{Piprime} with \eqref{defT10}, we deduce that each bivector of the family $(\vec{L},\vec{\Kp})$ is $\g$-simple. The frame of $\g$-simplicity is identified by $g$ as 
\begin{align}
N^I&=\L^I{}_J(g)t^J \label{FoD}\\\nn
&= \left(\sqrt{\frac{j^{2}+\rho^{2}}{k^{2}+\rho^{2}}}\cosh x,
\sqrt{\frac{j^{2}-k^{2}}{k^{2}+\rho^{2}}} \cosh x \frac{\r\vec p + j \vec q}{\sqrt{j^2+\r^2}} 
+ \frac{k}{j}\sqrt{\frac{j^{2}+\rho^{2}}{k^{2}+\rho^{2}}} \sinh x\, \vec{n}
-\frac{\rho}{j}\sqrt{\frac{j^{2}-k^{2}}{k^{2}+\rho^{2}}} \sinh x \, \vec{n}\times\frac{\r\vec p + j \vec q}{\sqrt{j^2+\r^2}} \right).
\end{align}
This give a two-parameter family of time-like normals, parametrized by the freedom of rotating $(\vec F, \vec n\times\vec F)$ to $(\vec{p},\vec{q})$ and by $x$. 
In other words, our construction gives a one-parameter family of bivectors, spanned by the rotational freedom, all of which are $\g$-simple in a frame that is determined by the spin $j$, the rotational parameter of the bivector, and a second free parameter $x$. These three parameters determine the frame uniquely. Notice that for non-minimal spin $j> k$, there is no choice of the other two parameters that would make the frame the canonical one. 
For $j=k$ on the other hand, the rotational parameter drops out, and the canonical frame is obtained setting $x=0$, which corresponds to $g=\Id$.

\noindent The homomorphism of Lorentz transformations from spinors to bivectors extends to this new family, 
\be
 \Pi'_\L= T^{(1,0)}(h) \Pi' = (k+i\r )\bra{\z}g^{-1}h^{-1}\vec{\s} hg\ket{\z}.
\ee
However the transformation of the electric and magnetic parts is now more complicated, with
\begin{align}
& \vec L\mapsto \vec L_\L=T^g_\g(h) \vec L, \qquad 
\vec \Kp\mapsto \vec \Kp_\L= 
T^g_{-1/\g}(h) T_{-1/\g}(g)^{-1} T_\g(g) \vec\Kp, \\
& T^g_\g(h) = \f kj R \left(\re(B) R_g \Big(\re(B_g) -\g  \im(B_g)\Big) - \im(B) R_g \Big(\g\re(B_g)+\im(B_g) \Big)\right).
\end{align}
In the last equation we used the polar decomposition for $g$, with notation $T^{(1,0)}(g)=R_gB_g$.
For lowest spins, $j=k$, the general map reduces to the previous one \eqref{Tgdef}, since in this case $g=\Id$ and $\vec\Kp=\vec K$.

\section{Coherent invariant tensors}

\subsubsection*{Invariant tensors in the orthonormal basis}

Consider the integrals of the representation matrices \eqref{Dh} with respect to the Haar measure $dh$, 
\be\label{uno}
{\cal I}^{(\r,k)}_{jmln}:= \int_{\SL(2,\C)} dh \prod_{a=1}^n D^{(\r_a,k_a)}_{j_am_al_an_a}(h).
\ee
These define tensors which are  invariant under global $\SL(2,\C)$ transformations. We should exclude from the definition the case $n=1$, which is divergent due to the non-compactness of the group. For $n=2$ the integral gives the (distributional) orthogonality relation of the matrices, and for $n=3$ the product of two Clebsch-Gordan coefficients. For $n\geq 4$, the result is the product of two generalized Clebsch-Gordan coefficients with $n$ legs, each of which can be decomposed in terms of $n-3$ basic coefficients.
See \cite{BarrettLorAsymp,Boosting,noiGen,Martin-Dussaud:2019ypf} for details on the recoupling theory of $\SL(2,\C)$.
When using graphical calculus, this integral is represented as a box traversed by $n$ strands. In reference to this, we will use the term strand in reference to a single value of $a$.

The exact evaluation of the integrals is hard, both analytically and numerically, because one is dealing with unbounded integrals of highly oscillating functions. Progress can be made using the Cartan decomposition which allows one to write the invariants as a one-dimensional unbounded integral, contracted with SU(2) Clebsch-Gordan coefficients \cite{Rashid70I,Boosting}. 
The unbounded integral can be further solved in terms of finite sums for $n=3$  \cite{Kerimov}, achieving an explicit analytic expression for the coefficients. Application of the same method for $n\geq4$ however simply trades the original integration for a new unbounded integration, this time over a virtual $\r$ label \cite{Boosting}. The decomposition is nonetheless very advantageous to improve numerical evaluations. A numerical code based on \cite{Boosting} was developed in \cite{Dona:2018nev}, improved in \cite{Gozzini-to-appear?}, and it is at the root of numerical evaluations of the Lorentzian EPRL spin foam model \cite{IoSU2asympt,Dona:2018pxq,noiLor}.

This said about the exact evaluation of the integrals, in this paper we are interested in  approximate evaluations in the limit of large quantum numbers.
This approximation is a semiclassical limit, in the sense that the spectra become denser and the relative uncertainties smaller.
The labels $m_a$ and $n_a$ are not very convenient to study this limit, because they only capture the projection of the angular momentum along the $\hat z$ axis, with maximal uncertainty along the transverse directions. Switching attention to the coherent  basis we have two clear advantages: a simpler formulation of the asymptotic problem, thanks to the factorization property of coherent states, and a sharper geometric interpretation.

\subsubsection*{Invariant tensors in the coherent basis}
The coherent invariants are a linear combination of \eqref{uno} weighted by the coherent states' coefficients, or equivalently group averages of the matrix elements in the coherent basis,
\be
{\cal I}^{(\r,k)}_{j \tl\z l \z} := \int_{\SL(2,\C)} dh \prod_{a=1}^n D^{(\r_a,k_a)}_{j_a\tl\z_al_a\z_a}(h).
\ee
For later convenience in the interpretation of the critical point equations, we exploit the invariance of the measure to rename the integration variables  $z_a^A\mapsto \bar z_a^A$ and $h_a\mapsto h^{\dagger-1}_a$. With this change of notation, and using \eqref{Dcs}, we have
\begin{align}\label{CTI}
{\cal I}^{(\r,k)}_{j \tl\z l \z}
=\left(\prod_{a=1}^n{\cal N}^{k_a}_{j_a l_a }\right)
\int_{\SL(2,\C)} dh \int_{(\C P^1)^n}\prod_{a=1}^n \f{d\m(z_a^A)}{\norm{z_a}^{2} \norm{h^{-1} z_a}^{2} }\, e^{S\left(h,z;\z,\tl\z\right)},
\end{align}
where
\begin{align}
\label{defS}
S(h,z;\z,\tl\z)&:=\sum_a \tl s(h^{\dagger-1},\bar z_a;\z_a,\tl\z_a) = \sum_a s(h , z_a; \z_a,\tl\z_a)\nn\\
& = \sum_a \log\f{ \bra{z_a}{\tl \z_a}]^{j_a-k_a} \bra{{\tl\z_a}} z_a\ra^{j_a+k_a} }{\norm{z_a}^{2(j_a+i\r_a)}} 
+ \log\f{ [\z_a \ket{ h^{-1}z_a}^{l_a-k_a} \bra{ h^{-1} z_a } \z_a \ra^{l_a+k_a}}{\norm{h^{-1} z_a}^{2(l_a-i\r_a)}}.
\end{align}
This type of integrals appear in Lorentzian spin foam models of quantum gravity. Following that literature, we will  refer to the parameters 
$(\r_a,k_a,j_a,l_a,\z_a, \tl\z_a)$ labelling the invariants as \emph{boundary data}. With the choice \eqref{plat}, the boundary spinors are identified uniquely with unit  $(\vec n_a, \vec{\tl n}_a)$. The formulas are valid also if we relax this choice and extend the boundary data to include a free global phase for each spinor, namely an arbitrary flag. Some care is due in this case since our definition of invariants refers to kets $\ket\z$ which are Perelomov coherent states with fixed phase convention, see \eqref{plat}, and not to arbitrary spinors. This has some implications for the symmetries of the action, which we discuss next.

\subsubsection*{Symmetries of the action}
The action is an even function of $h$, that is $S(-h)=S(h)$. It is invariant under rescalings 
\be\label{complexres}
z_a^A\mapsto\l_a z^A_a, \qquad \l_a\in\C,
\ee
as a consequence of the homogeneity of the representation functions.
It transforms inhomogeneously  under global rotations of either set of boundary vectors. This happens because it is defined in terms of SU(2) coherent states with a fixed choice of global phase, and rotating the boundary vectors induce a phase shift in the coherent states:
\be
\vec n_{a} \mapsto \vec n_a'=R \vec n_{a} \qquad \Rightarrow
\qquad \label{spinorrot}
\ket{\z_{a}} \mapsto \ket{\z_{a}^r} = e^{i \chi_{a} \rr}\ket{ \z_{a}}.
\ee 
Here $\rr$ is the $\SU(2)$ element corresponding to the rotation $R\in\SO(3)$, and $\chi_{a}=\chi_{a}(\vec n_{a},R)$  is related to the area of a spherical triangle \cite{Perelomov}. It follows that
\begin{equation}\label{actionshift}
 s(h,z_a;\tl{\z}^{\tl\rr}_a,\z^\rr_a) = s(\tl{\rr}^{\dagger}h\rr,\tl{\rr}^{\dagger}z_a;\tl{\z}_a,\z_a) +2i (l_a\chi_a - j_a \tl \chi_a). 
\end{equation}
As a consequence of \eqref{actionshift} and of the invariance of the measures of $h$ and $z_a$ under rotations, 
the coherent tensors transform by a phase under independent global rotations of the boundary data:
\be\label{Irotated}
{\cal I}^{(\r,k)}_{j \tl\z^{\tl \rr} l\z^\rr} = e^{2i\sum_{a}(l_{a}\chi_{a}-j_a\tl\chi_{a})} {\cal I}^{(\r,k)}_{j \tl\z l \z} .
\ee
The norm of the tensors is thus a function of rotational-invariant quantities only, namely of the scalar products
$\vec n_{a}\cdot \vec n_{b}$ and $\vec {\tl n}_{a}\cdot \vec {\tl n}_{b}$ between normals in the same set. If we extend the boundary data to include arbitrary spinors' flags, this only affects the global phase of the invariant, and not its norm.\footnote{Had we defined the coherent invariants in terms of arbitrary unit-norm spinors instead of SU(2) coherent states, as done e.g. in \cite{BarrettLorAsymp}, they would be exactly invariant under rotations of the boundary data. However, the map between boundary data and 3d normals would now depend on the rotation performed.}

\subsubsection*{A remark on phase conventions.}

Before proceeding with the study of the saddle point approximation of \eqref{CTI}, let us add a further remark on phases.
Naimark's basis \eqref{Naimark} is canonical only up to a phase factor $\exp\{i\Psi^\r_j\}$. This is set to zero in some literature including \cite{Ruhl,BarrettLorAsymp}, and we follow that convention here. It is a choice that leads to simpler formulas, but has the disadvantage of making the Clebsch-Gordan coefficients complex. An alternative phase convention leading to real Clebsch-Gordan coefficients is obtained multiplying \eqref{Naimark} by \cite{Kerimov,Boosting} 
\be\label{Psiphase}
e^{i\Psi^\r_{j}}  = (-1)^{-\f{j}2} \f{\G(j+i\r+1)}{|\G(j+i\r+1)|}.  
\ee
This is the phase convention used in the approach of \cite{Boosting} and in the numerical work \cite{Dona:2018pxq,Dona:2018nev,noiLor} implementing it. Its contribution to the semiclassical limit can be estimated straightforwardly and independently of the saddle point approximation, and will not be considered further in this paper.\footnote{The literature contains also an `intermediate' phase convention \cite{Strom,vongDuc1967,RashidBoost}, given by \eqref{Psiphase} without $(-1)^{-j/2}$. This choice simplifies the recursion relations satisfied by the  Clebsch-Gordan coefficients \cite{Rashid70I,Rashid70II}. The latter are now either real or purely imaginary.}

\section{Saddle point analysis}
The action \eqref{defS} depends linearly on the quantum numbers $(\r,k,j,l)$, thanks to the factorization property of the coherent states. This makes the integration amenable to a saddle point approximation for large quantum numbers. For the saddle point approximation to be meaningful, we fix $n\geq 3$ from now on.
Given the many parameters involved, there are many different possible asymptotics that one may wish to explore. 
In this paper we study the completely homogeneous limit, with all quantum numbers equally large, and finite differences among them:
\be
(\l\r_a,\l k_a,\l j_a,\l l_a), \qquad \l\rightarrow\infty.
\ee

A saddle point approximation of the integrals can be obtained looking for critical points at which the gradient of the action \eqref{defS} with respect to $h$ and $z^A_i$ vanishes.\footnote{It is possible to choose a section once and for all, and perform the calculation in adapted coordinates. This amounts to using a projective realization instead of the homogeneous realization of the principal series, and taking derivatives with respect to a single complex variable per strand, instead of the two of a spinor. We will however not do so: keeping the redundant spinorial variables has the advantage of giving simpler equations, as projective spaces are typically easier to manipulate using their homogeneous coordinates.} 
 In this paper we will restrict attention to the dominant saddles, namely those for which the real part of the action is maximal.
 The maximum is actually zero when all spins are lowest \cite{BarrettLorAsymp}, but not in general. As we will show below, the non-zero real part combines with the prefactors $\cal N$ eliminating any exponential behaviour.
The action is a complex function of complex variables, but has a simplifying feature: it is easy to identify the maximal value of its real part using the Cauchy-Schwartz inequality, and this simplifies the study of the gradient. 
Having an even action means also that the critical point equations are blind to the spin lift, and can be analysed using the vectorial representation $T^{(1,0)}(h)$ without any loss of generality. 

For configurations admitting critical points with non-singular Hessian matrix $H^\cc$, the standard formula for the leading order of the saddle point approximation gives 
\begin{align}\label{general}
{\cal I}^{(\l\r_a,\l k_a)}_{\l j_a \tl\z_a \l l_a \z_a} \simeq 
 \left(\prod_{a=1}^n{\cal N}^{\l k_a}_{\l j_a \l l_a }\right) \f{(2\pi)^{n+3}}{\l^{n+3}} 2
 \sum_{\cc} N^\cc e^{S\left(h^{\cc},z^{\cc};\z,\tl\z\right)},
\end{align}
where
\begin{equation}
N^\cc=\f{\Omega^\cc}{\norm{z^\cc_{a}}^{2} \norm{h^{\cc-1} z^\cc_{a}}^{2} } \f{1}{\sqrt{-\det H^\cc}},
\end{equation}
and $\Om^\cc$ is the value of the spinorial measure at the critical point.\footnote{We omit a numerical factor that come from the Haar measure. There is no preferred choice for this, because of the non-compact nature of the group manifold. } In this paper we will refrain from an explicit calculation of the Hessian (whose determinant can only be accessed numerically already for the lowest spin case \cite{noiLor}), but the explicit numerical tests reported below confirm that it is non-singular at the critical points we find.
The spare factor of 2 in \eqref{general} takes into account the fact that critical points come always in pairs because the  action is invariant under change of sign $h\to -h$. Accordingly,  the sum should be understood over  critical points with distinct value of the action. 
Notice that if we have two distinct critical points, the property \eqref{actionshift} guarantees that the relative phase between the two is invariant under global rotations of each set of boundary normals. This can be seen immediately since the shift term does not depend on $h$ nor $z_a$.
Therefore while the overall phase of the amplitude depends on the choice of phase for the coherent states as well as on the orientation of the boundary data, the relative phase between two distinct critical points does not.

\subsection{Spinorial gradient and the \magg equations}
The spinorial gradient can be studied strand by strand. To make the notation lighter, we remove the $a$ label in this section, and restore it in the next one when we will consider all strands at once.
Vanishing of the spinorial gradient gives
\begin{subequations}\label{gradz}\begin{align}
\left(\f{\p s}{\p \ket z}\right)^\dagger &= -(j+i\r)\f{\ket{z}}{\norm{z}^{2}}-(l-i\r)\f{h^{\dagger-1}h^{-1}\ket{z} }{\norm{h^{-1}z}^{2}} + (j+k)\f{\ket{\tl \z}}{\langle z \ket{\tl \z}}+(l-k)\f{h^{\dagger-1}\sket{\z}}{\langle h^{-1}z\sket{\z}} = 0,
\\ 
\f{\p s}{\p \bra z} &= -(j-i\r)\f{\ket{z}}{\norm{z}^{2}}-(l+i\r)\f{h^{\dagger-1}h^{-1}\ket{z} }{\norm{h^{-1}z}^{2}} + (j-k)\f{\sket{\tl \z}}{\langle z \sket{\tl \z}}+(l+k)\f{h^{\dagger-1}\ket{\z}}{\langle h^{-1}z\ket{\z}} = 0.
\end{align}\end{subequations}
These are four complex equations, but give only two complex conditions on the variables because of their homogeneity.
Among the possible critical points, we select only those that give maximal real part of the action.
Inspection of \eqref{defS} shows that this requires the four scalar products to have norm one. As a consequence, we can parametrize the dominant critical configurations by
\begin{subequations}\label{Re}\begin{align}\label{Re1}
& \f{\ket z}{\norm{z}}              = e^{i\tl\a}c_+^j \ket{\tl\z} + e^{i\tl\b} c_-^j \sket{\tl\z},  \\\label{Re2}
& \f{h^{-1}\ket z}{\norm{h^{-1} z}} = e^{i\a}   c_+^l \ket{\z   } + e^{i\b}    c_-^l \sket{\z   },
\end{align}\end{subequations}
where 
\be
c_{\pm}^j:=\sqrt{\frac{j\pm k}{2j}}.
\ee
Because of the rescaling invariance \eqref{complexres}, only three phases are relevant here: the relative ones $\a-\b$ and $\tl\a-\tl\b$, and the mixed one $\a+\b-\tl\a-\tl\b$. The relative phases depend on the global phase of the boundary spinors, and thus on their null flag. We refer to them as `flag' phases.
The mixed phase is independent of the spinors' flags and thus  dependent on the 3-vector data only. We refer to it as `vectorial' phase. 
Using a linear combination of the flag phases, the vectorial phase can be traded for $\a-\tl\a$. The latter  will be used for convenience in some of the following  algebraic manipulations. 

The condition \eqref{Re1} determines the section of $z^A$ in terms of the boundary data and the relative phase $\tl\a-\tl\b$. So to identify the critical data, one needs to solve for this phase and $h$, which in turn will require to solve for $\a-\b$. To isolate $h$, we proceed as follows. First, we combine \eqref{Re} in the single equation
\begin{subequations}\label{master}\begin{align}
\label{master1}
h\left(
c_{+}^l \ket{\z}      + e^{-i(\a-\b)}          c_{-}^l \sket{\z     } 
\right) &=\frac{\norm{z}}{\norm{h^{-1}z}}e^{-i(\a-\tl{\a})}\left(
c_{+}^j \ket{\tl{\z}} + e^{-i(\tl{\a}-\tl{\b})}c_{-}^j \sket{\tl{\z}} 
\right).
\end{align} 
Second, we simplify the gradient equations \eqref{gradz} inserting the condition of maximal real part \eqref{Re}. This gives a single independent spinorial equation,
\begin{align}
\label{master2}
h^{\dagger-1}\left(
 \left(l-i\r\right)                   c_+^l \ket{\z} 
-\left(l+i\r\right)e^{-i(\a-\b)}      c_-^l \sket{\z}
\right)&=\frac{\norm{h^{-1}z}}{\norm{z}}e^{-i(\a-\tl{\a})} \left(
 \left(j-i\r\right)                   c_+^j \ket{\tl\z} 
-\left(j+i\r\right)e^{-i(\tl\a-\tl\b)}c_-^j \sket{\tl\z}
\right).
\end{align}\end{subequations}
The two equations \eqref{master} can be rewritten in compact form as follows,
\begin{subequations}\label{masterg}\begin{align}
\label{eqphase0}
&  h  g_l \ket{\z} = \frac{\norm{z}}{\norm{h^{-1}z}}e^{-i(\a-\tl{\a})}         g_j            \ket{\tl{\z}}, \\
& (k-i\r)h^{\dagger-1}g_l^{\dagger-1} \ket{\z} = \frac{\norm{h^{-1}z}}{\norm{z}}e^{-i(\a-\tl{\a})} (k-i\r) g_j^{\dagger-1}\ket{\tl{\z}},
\end{align}\end{subequations}
in terms of two $\SL(2,\C)$ matrices
\begin{subequations}\begin{align}
& g_{j}= c_+^j    \ket{\tl\z}  \bra{\tl\z}
+e^{-i(\tl\a-\tl\b)} c_-^j \sket{\tl\z} \bra{\tl\z} 
+e^{ i(\tl\a-\tl\b)} c_-^j \frac{j-i\r}{k+i\r} \ket{\tl\z}  \sbra{\tl\z} 
+  c_+^j \frac{j+i\r}{k+i\r} \sket{\tl\z} \sbra{\tl\z}, \\
& g_{l}= c_+^l   \ket{\z}  \bra{\z}
+e^{-i(\a-\b)} c_-^l \sket{\z} \bra{\z} 
+e^{ i(\a-\b)} c_-^l \frac{l-i\r}{k+i\r} \ket{\z}  \sbra{\z} 
+  c_+^l \frac{l+i\r}{k+i\r} \sket{\z} \sbra{\z}.
\end{align}\end{subequations}
This rewriting will now allow us to obtain a simple geometric interpretation of the critical point equations. 
Contracting the two equations with the insertion of $\vec \s$, we obtain
\begin{align}\label{maggg}
\bra{ \z} g_l^{-1} h^{-1}\vec\s h g_l \ket{\z} 
= \bra{ \tl\z} g_j^{-1}\vec\s g_j \ket{\tl \z}. 
\end{align}
This equation can be understood with the help of the homomorphism between Lorentz transformations on spinors and bivectors, discussed in Section~\ref{sec:gsimple}. Using \eqref{defT10}, we see that \eqref{maggg} is equivalent to
\begin{align}\label{magg}
& T^{(1,0)}(hg_l) (k+i\r)\vec{n} 
= T^{(1,0)}(g_j) (k+i\r)\vec{\tl n}.
\end{align}
This is our key equation.

A few remarks are in order. First of all, \eqref{magg} is enough to determine all the variables  for any $n\geq 3$. We have $4n$ real equations, and $6+2n$ variables, 6 in $h$ plus $2$ components for each $z_a$, or equivalently the flag phases in our parametrization \eqref{Re}. 
Strictly speaking, we will solve in this way for $T^{(1,0)}(h)$. This determines $h$ up to a sign, but since the action is even this is just a redundancy in the multiplicity of critical points.
The vectorial phase has dropped out from \eqref{magg}. It can be determined once the explicit solution for $h$ and $z_a^A$ is known, and we will do so below.
Most importantly, \eqref{magg} endows the critical point equations with a clear geometric interpretation, which guides us in searching for solutions. The boundary data must be such that they describe two sets of $\g$-simple bivectors, in frames related to the spins $j_a$ and $l_a$, that can be mapped into one another by a single Lorentz transformation.

To see this construction more explicitly, we expand \eqref{magg} using the explicit form of $g_j$ and $g_l$. This gives
\begin{subequations}\begin{align}\label{TBB}
&T^{(1,0)}(h) 
\left(
\left(l+i \f{\r k}{l}\right)\vec{n} + i\frac{\sqrt{l^2-k^2}}{l} (l\vec p - \r \vec q\, )
\right)  
= 
\left(j+i \f{\r k}{j}\right)\vec{\tl n} +i \frac{\sqrt{j^2-k^2}}{j} (j\vec{\tl{p}} - \r \vec{\tl{q}}\, ),
\end{align}
with
\begin{align}\label{defpq}
& \vect{\vec p}{\vec q} = \mat{\cos(\a-\b)}{\sin(\a-\b)}{-\sin(\a-\b)}{\cos(\a-\b)} \vect{\vec F }{\vec n\times\vec F},\\
& \vect{\vec {\tl p}}{\vec {\tl q}} = \mat{\cos(\tl\a-\tl\b)}{\sin(\tl\a-\tl\b)}{-\sin(\tl\a-\tl\b)}{\cos(\tl\a-\tl\b)} \vect{\vec {\tl F }}{\vec{\tl n}\times\vec {\tl F}}.
\end{align}
\end{subequations}
We see that the critical point equations have naturally introduced the boosted $\g$-simple bivectors discussed in Section~\ref{sec:gsimple},
with the flag phases playing the role of the free parameter of rotations with respect to the basis \eqref{evsigma}.
In fact, one can check that $g_j$ is precisely of the form \eqref{4screw}, with $(\vec p,\vec q)$ related to the flag phases by \eqref{defpq}, and the second four-screw being in the $\vec{\tl n}$ direction with complex angle $x+iy=\log \frac{k+i\r}{j+ i \r}$.
Therefore, $T^{(1,0)}(g_j) (k+i\r) \vec{\tl n}$ is precisely (the self-dual part of) a $\g$-simple bivector. 

Solving the critical point equations thus amounts to finding the frame of $\g$-simplicity, via the solution for the flag phases, and determining the Lorentz transformation relating the two sets of bivectors. 
We will discuss how to do this in the next Section, after completing the study of the gradient below.
Notice that by construction, the critical point equations determine the frame of $\g$-simplicity on each strand $a$ only up to a four-screw in the direction of $\vec{\tl n}_a$.
We expect that this freedom can be used to relate the strand-dependent frames on each set to a global frame, and then the critical group element is the one transforming these two global frames. This expectation can be explicitly verified in the half-lowest case (it is trivial in the lowest case).

\subsection{Gradient in the group variable and the closure condition}
To compute the gradient in the group variables it is simplest to look at variations of the action with respect to variation $\d h$. The latter can be expressed in terms of algebra elements via
\be\label{dXdef}
\d X :=  h^{-1} \d h = \f i2\vec a\cdot\vec\s+\f 12\vec b\cdot\vec\s\in\sl(2,\C).
\ee 
This gives
\begin{align}
{\d_h S} &= 
\sum_{a}  (l_a-k_a)  \f{\sbra{\z_a}\d X h^{-1}\ket{z_a} }{\sbra{\z_a}h^{-1}\ket{z_a}} +
 (l_a+k_a)  \f{\bra{z_a}h^{\dagger-1}\d X^\dagger \ket{\z_a} }{\bra{z_a}h^{\dagger-1}\ket{\z_a}} 
-(l_a+i\r_a)\f{\bra{z_a}h^{\dagger-1}(X + X^{\dagger})h^{\dagger-1} \ket{z_a}}{\norm{h^{-1}z_a}^2} \nn\\
&= - i {\vec a} \cdot \sum_a l_a\vec n_a 
- i\vec b \cdot \sum_a \f{\r_ak_a}{l_a}\vec n_a+\f{\sqrt{l^2_a-k^2_a}}{l_a} (j\vec p_a-\r\vec q_a) 
= i \vec{a} \cdot \sum_{a} \vec{L}_a + i \vec{b} \cdot \sum_{a} \vec{\Kp_a}.
\label{gradhS}
\end{align}
Here we used the  critical point equation \eqref{Re} and the definition \eqref{defpq} in the second equality, and the definition \eqref{KFF}  of $\vec\Kp$ in the last. Only one set of data appears, determined by the choice of insertion of $h$ in the $F$ functions. 
Again, we see that the boosted $\g$-simple bivectors appear naturally from the critical point equations.
The vanishing of this gradient  implies 
\be
\label{closure}
\sum_a\vec L_a = \sum_a\vec \Kp_a = 0 \quad \Leftrightarrow \quad \sum_a\vec \Pi_a' = 0.
\ee
This is a \emph{closure conditions} on both the electric and magnetic parts of the $\g$simple bivectors $(L_a,\Kp_a)$.
Since we are free to move $h$ in the action from one set of scalar products to the other without changing the integral, the same closure conditions must hold also for the with the tilded bivectors,
\be
\sum_a\vec {\tl L}_a = \sum_a\vec {\tl \Kp}_a = 0 \quad \Leftrightarrow \quad \sum_a\vec {\tl \Pi}_a' = 0.
\ee

Closure conditions appear also in SU(2) coherent invariant tensors \cite{LS,noiGen}, and are the semiclassical version of the quantum invariance \cite{Kapovich,IoPoly}. Their solution is well-known in geometric terms.
Each real closure condition defines a (generically bent) polygon in $\R^3$. But for $n\geq4$ and non-coplanar normals, it identifies also a \emph{unique} convex polyhedron \cite{minkowski1897allgemeine,IoPoly}, with the vectors determining the area and the normal direction of each face. The case of non-coplanar data and their 3d polyhedral interpretation is the one most directly relevant to LQG.

As critical point equations, the closure conditions do not involve the integration variables $h$ and $z_a$. 
They are directly restrictions on the boundary data, which must describe sets of polygons or polyhedra in $\R^3$. This can imply restrictions on the irrep labels $(\r_a,k_a)$, depending on the normals. 
The value of the integration variables is thus determined entirely by the vanishing of the gradient in the $z_a$ variables, and the associated equations \eqref{magg} derived earlier.
By providing a global frame for each set and a Lorentz transformation, the critical point equations allows us to embed $\R^3$ space in Minkowski space, and to visualize $h$ as the actual Lorentz transformation mapping one polygon or polyhedron to the other, with the action defined via the bivector map.

\subsection{Action at the critical points}
Using \eqref{Re}, the strand action at the critical point reads\footnote{This formula is valid if $\a+\b\in[-\pi,\pi)$. There is an additional term $ik\pi$ if $\a+\b\in[\pi,2\pi)$, which follows from taking into account the appropriate branch of the complex logarithm. This restores the even-ness of the action, if we remark \label{pigna}
from the parametrization \eqref{Re} that the phases $\a$ and $\b$ record the overall sign of $h$.}
\begin{align}
\label{critaction}
s^\cc &= \frac{1}{2} \log \f{(j+k)^{j+k}(j-k)^{j-k}}{(2j)^{2j}}+ \frac{1}{2} \log \f{(l+k)^{l+k}(l-k)^{l-k}}{(2l)^{2l}} 
\nn\\&\quad + i \Big(\r \log \f{\norm{z}^2}{\norm{h^{-1}z}^2} - k(\a+\b-\tl{\a}-\tl\b)+  j (\tl{\a}-\tl{\b})- l (\a-\b)\Big).
\end{align}
As anticipated earlier, the real part vanishes only in the case of lowest spins, but is in general positive.
To evaluate the imaginary part, it is convenient to derive equations for the vectorial phase and ratio of norms as functions of $h$ and of the flag phases. 
These can be simply obtained from scalar products of \eqref{master}. One option is
\begin{align}
\label{normratio}
& \bra{\z} g_l^\dagger h^{\dagger}h g_l \ket{\z} =\frac{\norm{z}^2}{\norm{h^{-1}z}^2}, \\
\label{mixedphase}
&\bra{\z} g_l^\dagger h^{\dagger}h g_l \sket{\z} 
=\frac{\sqrt{j^2 - k^2}}{k+i \rho} e^{i(\tl\a-\tl\b)} e^{2i(\a-\tl\a)},
\end{align}
which has the advantage of highlighting the role of the pure boosts. Another option is
\be\label{lowestphase}
\bra{\tl{\z}}hg_{l}\ket{\z} =\f{\norm{z}}{\norm{h^{-1}z}}e^{-i(\a-\tl{\a})}c_+^j.
\ee
This is useful for instance when both SU(2) spins are lowest, configuration for which \eqref{mixedphase} is an identity.

\section{Solutions and asymptotic formulas}\label{SecAsymp}

The main technical step to solve the boosted orientation equations \eqref{magg} is isolating the flag phases from $h$.
This can be done simply taking real and imaginary parts, if at least one SU(2) spin $j$ or $l$ at its lowest value $k$.
When both spins are general, we were not able to fully disentangle the equations. 
In the light of this difficulty, and in order to simplify the discussion of the geometric interpretation of the solutions, we will split the analysis of the solutions in three cases, in growing order of complexity. 

\subsection{Lowest case}\label{SecL}

By lowest case we mean the configurations with both sets of SU(2) spins at their minimal value,
\be
j_a = l_a = k_a >0.
\ee
In this case, the asymptotic analysis of the invariants is a straightforward adaptation of the results already known in the literature.
Consider first the closure condition. Taking $j_a=k_a$, we have
$\vec\Kp_a=\vec K_a =\g_a \vec L_a$, and  \eqref{closure} simplifies to 
\be\label{closcond}
\sum_a k_a \vec n_a = 0, \qquad \sum_a \r_a \vec n_a = 0.
\ee
Taking $l_a=k_a$, and using the freedom to move $h$ in the integrand, we obtain the same result for the tilded data.
With lowest spins, the vectors $(\vec p,\vec q)$ have dropped out of the analysis, and both closures in \eqref{closcond} involve the same normals. They thus describe two conformal polygons, sharing the same angles but with different lengths. We can also see explicitly examples of the restrictions on the irrep labels $(\r_a,k_a)$. 
For instance for singular configurations with all vectors aligned, the sum of the quantum labels $k_a$ must be even.
For $n=3$, and for $n=4$ with non-coplanar vectors $\vec n_a$, the closure conditions can only be satisfied if
\be\label{g!}
\g_a=\g \quad \forall a.
\ee
We then have a single polygon or polyhedron, up to a global rescaling.
For 4 coplanar vectors, or $n\geq 5$, there exist solutions without the labels being all proportional, if a subset of normals independently closes. 
In this case different $\g_a$ are allowed, and we have two different polygons or polyhedra associated with the same set of normals $\vec n_a$, and different values of the lengths or areas. 

Next we move to the spinorial gradient. We have $c_-^{j}=c_-^l=0$, and the maximality conditions \eqref{Re} contain only the first term on the right-hand side. There is a single relevant phase, the vectorial phase $\a_a-\tl\a_a$. The matrices $g_{j_a}$ and $g_{l_a}$ reduce to the identity, and \eqref{master}  can be recognized as the critical point equations that appear in \cite{BarrettLorAsymp}, with the simplification of having a single group element and no graph structure. We can write the vectorial critical point equations \eqref{magg} as
\be
\label{magglow}
T^{(1,0)}(h) (k_a+i\r_a)\vec{n}_a = (k_a+i\r_a)\vec{\tl n}_a.
\ee
We remark that in this simpler case, we can use a map from the boundary data to self-dual bivectors with vanishing magnetic part, instead of the $\g$-simple bivectors. This alternative map is used in \cite{BarrettLorAsymp,Dona:2020yao}.

This equation involves only $h$ as variable, and the solution is easy.
Using the polar decomposition $h= \rr b$ and squaring the real part of \eqref{magglow} we find 
\begin{equation}
(T_\g (h) \vec{n}_a)\cdot (T_\g (h) \vec{n}_a) = \vec{\tl n}_a \cdot \vec{\tl n}_a= 1 \quad \Rightarrow \quad  \norm{\vec{u}\times \vec{n}_a}^2\sinh^2 \eta= 0,
\end{equation}
where $\vec u$ is the direction of the boost and $\eta$ its rapidity.
This  is solved by either $\eta =0$, in which case the critical group element is a rotation; or by $\vec{u}\times \vec{n}_a=0$, 
and the critical group element is a four-screw in direction $\vec{n}_a$.
At this point we must distinguish two cases: singular configurations, for which all vectors of one set are aligned, and non-singular configurations. For non-singular configurations, it is not possible to have $\vec{u}\times \vec{n}_a=0$ for all $a$. Therefore, the solution of \eqref{magglow} is a pure rotation, and corresponds geometrically to the rotation between the two polygons or polyhedra described by the boundary data. Such rotation is uniquely determined given the boundary data.  We refer to Appendix~\ref{AppA} for an explicit proof of the uniqueness, and an algorithm to compute $h^\cc$ explicitly. 
In the singular case, this unique rotation can be multiplied by an arbitrary four-screw in the  direction of the alignment. 

We thus have two critical points $h^\cc=\pm \rr^\cc$ for non-singular configurations (which we call non-distinct because they give the same contribution to the saddle point approximation), and two times a 2d infinite set for singular configurations.

\subsubsection*{Asymptotics}
The numerical prefactor on each strand is 
\be\label{defcNm}
{\cal N}^k_{kk}=\frac{d_k}{\pi}.
\ee
The action \eqref{critaction} for lowest spins is purely imaginary. For non-singular configurations, $h^\cc\pm \rr^\cc\in\SU(2)$ and thus the norm ratio is one, so the action has a single contribution from the vectorial phase,
\begin{align}
S^{\cc}=i\Phi^{\cc}:= 2i \sum_a k_a(\a_a-\tl\a_a)^\cc.
\end{align}
The phase can be read from \eqref{lowestphase}, which gives $
\exp\{-i(\a_a-\tl{\a}_a)^\cc\} = \pm\bra{\tl\z_a} r^\cc\ket{\z_a}.$ The explicit form of the critical group element can be computed from the algorithm given in Appendix~\ref{AppA},
and can be always set to the identity appropriately rotating the boundary data.
Assuming non-degeneracy of the Hessian, the leading order is 
\begin{align}\label{LOms}
{\cal I}^{(\l\r_a,\l k_a)}_{\l k_a \tl\z_a \l k_a \z_a} 
&=  \f{2^{2n+4}\pi^3}{\l^{3}} \left(\prod_{a=1}^n k_a \right)
N^\cc e^{i\Phi^{\cc}} +o(\l^{-3}).
\end{align}
The scaling of the power law of critical configurations is independent of $n$.

For singular configurations, there will be two flat directions in the Hessian corresponding to rotations and boosts around and along the axis of symmetry. 
Hence the overall scaling is $\l^{-2}$ instead of $\l^{-3}$, again independently of $n$.

\subsection{Half-lowest case}\label{SecHL}
By half-lowest case we mean configurations with only one set SU(2) spins is at their minimal value,
\be
j_a > k_a > 0, \qquad l_a = k_a >0.
\ee
The choice of $l_a$ minimal simplifies the treatment since those are the scalar products involving $h$. The other option can be easily obtained by transposition of matrices.

With this configuration, the untilded bivectors are $\g$-simple in the canonical frame, and the tilded ones in the $j_a$-dependent boosted frame, 
\be
\vec{\Pi}_a = (k_a + i \rho_a) \vec{n}_a, \qquad \vec{\tl \Pi}_a = (k_a + i \rho_a) T^{(1,0)}(g_{j_a})\vec{\tl n}_a.
\ee 
The closure conditions require the sums of these complex vectors to vanish. For the untilded data, this gives \eqref{closcond} discussed earier. For the tilded data, we obtain
\be\label{closcond2}
\sum_a j_a \vec{\tl n}_a = 0, \qquad \sum_a \f{\r_ak_a}{j_a}\vec {\tl n}_a+\f{\sqrt{j^2_a-k^2_a}}{j_a}(j_a\vec {\tl p}_a-\r_a \vec {\tl q}_a)=0.
\ee
In this case the two electric and magnetic polygons are not conformal anymore. 
The electric polygons and the Lorentz transformation relating them provide the referential geometric interpretation of the critical point equations, whereas the different geometry of the magnetic polygons capture the presence of a boosted frame.
To study the Lorentz transformation relating them, we turn our attention to the spinorial gradient. Only $c_-^l$ vanishes, and we there are two relevant phases: the flag phases $\tl{\a}_a-\tl{\b}_a$ and the vectorial phases  $\a_a-\tl{\a}_a$, or equivalently $2\a_a-\tl\a_a-\tl\b_a$. The matrices $g_{l_a}$ are  the identity and the \magg equations reduce to
\begin{equation}
 T^{(1,0)}(h) \left(k_a+i \r_a\right) \vec{n}_a 
= T^{(1,0)}(g_{j_a}) \left(k_a+i \r_a\right) \vec{\tl n}_a=
\left(j_a+i \f{\r_a k_a}{j_a}\right)\vec{\tl{n}}_a - i\frac{\sqrt{j_a^2-k_a^2}}{j} (-j_a\vec{\tl{p}}_a + \r_a \vec{\tl{q}}_a).
\end{equation}
We see by inspection that taking  real and imaginary parts we can isolate equations for $h$ and for the flag phases. Furthermore, it has the effect of introducing the transformation matrix $T_\g(h)$ that we studied in Section~\eqref{sec:gsimple} in relation to $\g$-simple bivectors. A simple algebra in fact gives for the real and imaginary parts respectively,
\begin{subequations}
\label{Tg}
\begin{align}
k_a T_{\g_a}\left(h\right)\vec{n}_a&= j_a \vec{\tilde{n}}_a, \label{Tg1}\\
\rho_a T_{-1/{\g_a}}\left(h\right)\vec{n}_a&=\frac{\rho_a k_a}{j_a}\vec{\tilde{n}}_a
+\sqrt{j^{2}_a-k^{2}_a}\left(\vec{\tilde{p}}_{a}- \frac{\rho_a}{j_a} \, \vec{\tilde{q}}_{a}\right). \label{Tg2}
\end{align}
\end{subequations}
The real parts can be used to determine $h$, without knowledge of the flag phase required. In a sense, we are using the canonical frame provided by the data with lowest spins as reference. Using the polar decomposition $h= \rr b$ and taking the square of \eqref{Tg1} we obtain
\begin{equation}\label{Marco}
\sinh^2 \eta= \frac{j_a^{2}-k_a^2}{k_a^2\left(1+\g_a^{2}\right)\left(1-\left(\vec{n}_a\cdot\vec{u}\right)^{2}\right)}.
\end{equation}
For a single strand, this equation has always two solutions for the rapidity $\eta$. 
To be admissible on the full set of strands, the value of the right-hand side must be independent of $a$. This is a non-trivial condition on the labels, and it is part of the requirement already discussed in general below \eqref{magg} that the boundary data must described $\g$-simple bivectors that can be Lorentz transformed into one another.
We remark that the solution for the rapidity is invariant under rotations of the boundary data, since the direction $\vec u$ changes only under a rotation of the untilded normals, by choice of order of the polar decomposition used.

To determine the direction $\vec{u}$ and the rotation $\rr$ we need to use the equations on multiple strands.
We proceed in two steps. First, we exploit the covariance of the amplitude \eqref{Irotated} to rotate the boundary data so that $\rr=\Id$. 
The direction $\vec{u}$ can then be determined as in the lowest case, taking scalar products of $\vec{u}$ with \eqref{Tg1} for distinct $a$. This gives a unique solution $h^\cc=b^\cc$, and the proof and algorithm to reconstruct the corresponding pure boost are again referred to Appendix~\ref{AppA}. 
Since the pure boost solution is unique, if there are additional distinct solutions, they must include rotations. We now prove that there is indeed a second distinct solution for non-singular data, but only if \eqref{g!} holds, namely the irrep labels are all proportional to one another.

\subsubsection*{Multiplicity and geometry of critical configurations}
\label{multi}
We denote $h_1=b_1$ the unique pure boost solution, which can be constructed suitably rotating the boundary data. 
If a second solution $h_2 = \rr_2 b_2$ exists, it must satisfy \eqref{Tg1}. If the normals are non-coplanar, this implies that $T_\g(h_1)=T_\g(h_2)$, or $T_\g(b_1)=R_2T_\g(b_2)$ in terms of the polar decomposition. From this we derive $R_2 = T_\g(b_1)T^{-1}_\g(b_2)$ and $T_\g(b_1)^{\scr T} T_\g(b_1) = T_\g(b_2)^{\scr T}T_\g(b_2)$.
If we now recall that $T_\g$ is the composition of a rotation and a dilation, see \eqref{decomposition}, the last equation implies $D^2_\g(b_1) = D^2_\g(b_2)$. Explicitly,
\begin{equation}
\label{eq:mult}
d_\g^2(\eta_1)\mathds{1}+(1-d_\g^2(\eta_1))\vec{u}_1\otimes\vec{u}_1 =d_\g^2(\eta_2)\mathbb{I}+(1-d_\g^2(\eta_2))\vec{u}_2\otimes\vec{u}_2.
\end{equation}
Taking the trace of this equation we find $d_\g^2(\eta_1)=d_\g^2(\eta_2)$, implying that $\eta_2=\pm \eta_1$. Inserting this condition back in \eqref{eq:mult} we find that $\vec{u}_2=\pm\vec{u}_1$. The boost part of the candidate second solution have direction $\vec u_1$ and rapidity $\pm \eta_1$.
The first option gives $b_2=b_1$, and therefore $R_2=\Id$. This is just the initial solution.
The second option gives $b_2=b_1^{-1}$, and  $R_2 = T_\g(b_1)T^{-1}_\g(b_1^{-1})= R_{\vec u}(2\psi)$ with angle $2\psi(\g,\eta)$ defined in \eqref{psiangle}. Since $\eta$ is strand independent by requirement of the existence of the first solution, the existence of the second solution requires also $\psi$ to be $a$-independent. Inspection of \eqref{psiangle} shows that this can only occur if $\g_a=\g$ for all $a$.

This analysis proves that when there is a critical point in the half-lowest case, there is also a distinct one differing by the $\psi$ rotation. 
Restoring the possibility of arbitrary orientations of the two sets of data, the general form of the solutions is
\begin{equation}\label{twosols}
h^\cc_1 = \pm r^\cc e^{ {\eta} \vec{u} \cdot \f{\vec{\sigma}}2},\qquad 
h^\cc_2 = \pm r^\cc e^{ i \psi \vec{u} \cdot \vec{\sigma}} e^{-{\eta} \vec{u} \cdot \f{\vec{\sigma}}3}. 
\end{equation}
Here $r^\cc=\Id$ if the data are chosen such that the first solution is a pure boost, and in general can de computed uniquely from the `deboosted' boundary data with the procedure described in Appendix~\eqref{AppA}. The freedom of rotating the boundary data can also be used to write the two solutions in a more symmetric form, 
\be\label{hsolhm}
 h^{(\pm)} = \pm \rr '{}^\cc e^{ \pm (i \psi +\eta)\vec{u} \cdot \f{\vec{\sigma}}2}.
\ee
Namely a four-screw and its inverse in the direction $\vec{u}$, up to rotations.
 
\begin{figure}[h]
\centering
\includegraphics[width=10cm]{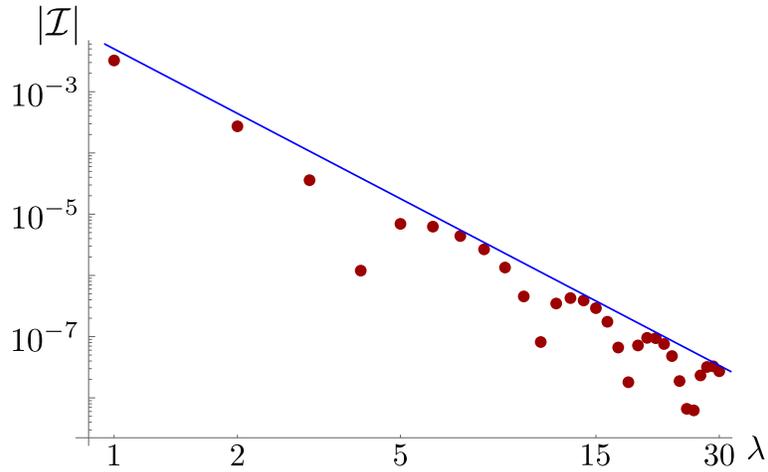}
\caption{\label{Fig3} \small{\emph{Numerical evaluation of the absolute value of the coherent invariant with $n=3$, for boundary data corresponding to two an equilateral triangle with $l_a=k_a=1$, boosted to an isosceles triangle with $l_a=(1,2,2)$, with $\gamma=1.2$. The plot is consistent with the expected $\l^{-\frac{7}{2}}$ decay -- see asymptotic formula at the end of this Section, added with a ad hoc coefficient to help the eye, and with the presence of  distinct critical points, because of the oscillations. 
Without explicitly computing the Hessian determinant to compare quantitatively the slope coefficents, the numerics alone cannot tell us the precise number of distinct configurations. The expectation is that there are two distinct critical points, just like for non-coplanar configurations with $n\geq 4$. 
}}
} 
\end{figure}

The proof that there are two distinct solutions assumes non-coplanarity, therefore the case $n=3$ would be excluded by the closure condition.
However, it is possible to adapt the above argument to take place with  two-by-two matrices restricted to the plane of the normals,
and a moment of reflection suggests that the conclusions are unchanged.
We thus expect that the existence of two distinct critical points occurs also for $n=3$.
We refrain from undertaking a detailed investigation of this case, and provide instead a numerical test that shows the presence of distinct critical points  (and finite in number, so not to change the power law), see Fig.~\ref{Fig3}.
On the other hand, the assumption of non-singular data remains necessary. If the normals are all aligned, we expect infinitely many solutions corresponding to arbitrary four screws along the direction of alinement, like in the lowest spin case.

\subsubsection*{Determination of the flag phases}

To finish solving the critical point equations, we use \eqref{Tg2} to determine the flag phases. First of all, we subtract it to \eqref{Tg1} multiplied by $\g_a$.
Using \eqref{Tgdef} and \eqref{Marco} we find
\begin{align}\label{HMflag1}
\sgn(\eta) \sqrt{1+\g_a^{2}}\, R \f{\vec{u}\times\vec{n}_a}{\norm{\vec{u}\times\vec{n}_a}}&=
 \g_a\f{\sqrt{j^{2}_a-k^{2}_a}}{j_a}\vec{\tl{n}}_a - \vec{\tl p}_a + \g_a\f{k_a}{j_a} \vec{\tl q}_a.
\end{align}
This equation should be understood to be valid at the critical point only, but we omitted the superscript $\cc$ in $\vec{u}$, $\eta$ and $R$ to keep the notation light.
There are two independent components in \eqref{HMflag1}, that can be extracted via scalar projections. 
Taking the scalar product with $\vec{\tl{n}}_a$ we obtain a component independent of the flag phases,
\be\label{HMflag2}
\sgn(\eta) \sqrt{1+\g_a^{2}}\, \tl n_a \cdot R \f{\vec{u}\times\vec{n}_a}{\norm{\vec{u}\times\vec{n}_a}}=
 \g_a\f{\sqrt{j^{2}_a-k^{2}_a}}{j_a},
\ee
which is one of the conditions that $h$ has to satisfy to boost one set of bivectors in the other, and therefore an identity at this level.
Taking the scalar product of this equation with $\vec{\tl{F}}_a+i \vec{\tl{n}}_a\times\vec{\tl{F}}_a$ we obtain an equation for the flag phases, 
\begin{equation}
\label{critflag}
e^{i(\tl\a_a-\tl\b_a)^\cc}= \sgn(\eta) e^{i\arctan\f{\g_ak_a}{j_a}} 	\f{j_a \sqrt{1+\g_a^{2}} }{\sqrt{j_a^2+ \g_a^{2}k_a^2} } \left(\vec{\tl{F}}_a+i\vec{\tl{n}}_a\times\vec{\tl{F}}_a\right)\cdot R \f{\vec{u}\times\vec{n}_a}{\norm{\vec{u}\times\vec{n}_a}}.
\end{equation}
The absolute value of the right-hand side is not manifestly one, but this can be explicitly checked on-shell of the critical point equations.\footnote{The squared absolute value of the scalar product in the right-hand side of \eqref{critflag} is
\be\nn
\left(\vec{\tl{F}}_a \cdot\f{\vec{u}\times\vec{n}_a}{\norm{\vec{u}\times\vec{n}_a}}\right)^2+
\left(\vec{\tl{n}}_a\times\vec{\tl{F}}_a\cdot\f{\vec{u}\times\vec{n}_a}{\norm{\vec{u}\times\vec{n}_a}}\right)^2 = 
1- \left(\vec{\tl{n}}_a \cdot\f{\vec{u}\times\vec{n}_a}{\norm{\vec{u}\times\vec{n}_a}}\right)^2= 1- \f{\g_a^2}{1+\g_a^2}\f{j_a^2-k_a^2}{j_a^2}.
\ee
Here we used the fact that ($\vec{\tl n}_a$, $\vec{\tl F}_a$, and $\vec{\tl n}_a\times \vec{\tl F}_a$) is an orthonormal basis, and 
 \eqref{HMflag2} in the last step.}
We can also check the invariance of the critical value of the flag phases under independent rotations of both sets of vectors. The mechanism is different for the two sets, because of the order of the polar decomposition we work with, namely $h=br$. Hence a rotation of the tilded normals is immediately absorbed in the critical rotation $R$, and the invariance of \eqref{critflag} is manifest. A rotation of the untilded normals has to be first commuted with the boost, and therefore also changes the critical boost direction $\vec u$. We see by inspection that \eqref{critflag} is invariant under this operation as well.

Using \eqref{Marco} and \eqref{HMflag1} the second closure condition $\sum \vec \Kp_a$ becomes an identity on-shell of the other three closures \eqref{closcond} and the first of \eqref{closcond2}. In other words, once the flag phases are on-shell, their part of closure is automatically satisfied and gives no further constraint.

\subsubsection*{The global frames}

The frame for each bivector with lowest spin is the canonical one, and this determines already a global frame $t^I$ for the first set of boundary data.
The critical Lorentz transformation maps this frame to a global frame for the second set, 
\begin{equation}\label{Ncrit}
N^\cc:=\L(h^\cc)t= (\cosh \eta , \sinh \eta \,R \vec{u}).
\end{equation}
The $\psi$-rotation drops out so if there are two distinct critical points, they give the same global frame.
The frame of $\g$-simplicity for each bivector in the second set is given by \eqref{FoD} with the critical values of the flag phases computed just above, and 
\be
x_a=\f12\log\f{k_a^2+\r_a^2}{j_a^2+\r_a^2}.
\ee 
Recall that the strand-dependent frames are defined only up to four-screws in the $\vec {\tl n}_a$ direction. We can use this freedom to match them to the global one:
\begin{align}\label{anvedi}
N^{\cc} &= \L^\cc_a N_a, \\
\L^\cc_a&:=\L\left(\exp\left((w_a-x_a) \vec{\tl n}_a\cdot\f{\vec\s}2\right)\right), \qquad \tanh w_a =\f{j_{a}}{k_{a}}\sqrt{j_{a}^{2}-k_{a}^{2}}\frac{\vec{u}\cdot\vec{\tl n}_{a}}{j_{a}\vec{u}\cdot\vec{\tl q}_{a}+\r_a\vec{u}\cdot\vec{\tl p}_{a}}.
\end{align}
This equality requires that all variables are at their critical value. 

Coming back to the bivectors $B_a = (\Id-\g\star) J_a$, the same four-screws make them all orthogonal to the same normal:
\begin{equation}
( \L_a^\cc N^\cc)_I B_a^{IJ} = 0.
\end{equation}
This equation defines the Minkowskian embedding of the boosted polyhedron.

\subsubsection*{Asymptotics}
When the SU(2) spins are not at their lowest value, the critical action picks up a non-vanishing real part, 
\begin{align}
\re(s^\cc_a) = 
\frac{1}{2} \log \f{(j_a+k_a)^{j_a+k_a}(j_a-k_a)^{j_a-k_a}}{(2j_a)^{2j_a}}.
\end{align}
In the saddle point approximation of the integrals, this combines with the prefactors ${\cal N}^{k_a}_{j_a k_a}$ given by \eqref{defcN}, to give
\begin{align}
& \frac{\sqrt{2\l j_a+1}\sqrt{2\l k_a+1}}{\pi}\left(\begin{array}{c}2\l j_a \\\l j_a+\l k_a \end{array}\right)^{\frac 1 2} 
\left(\frac{(\l j_a+\l k_a)^{\l j_a+\l k_a}(\l j_a-\l k_a)^{\l j_a-\l k_a}}{(2\l j_a)^{2\l j_a}}\right)^{\frac 1 2}
\nn\\& \qquad \approx \lambda^{\f 3 4} \;\frac{2\sqrt{j_a k_a}}{\pi}\left(\frac{j_a/\pi}{j_a^2-k_a^2}\right)^{\frac 1 4}
\label{Stirling}\end{align}
using the Stirling approximation.

The value of the imaginary part of the action depends on the configuration of the normals at the critical point. 
From \eqref{normratio} and the polar decomposition we have
\begin{equation}
\label{normratio2}
\f{\norm {z_a^\cc}^2}{\norm{h^{\cc -1}z_a^\cc}^2} = \bra{\z_a} h^{\cc \dagger}h^\cc \ket{\z_a} = \cosh \eta -\vec{u}\cdot\vec{n}_a \sinh \eta.
\end{equation}
The solution for the rapidity is given by \eqref{Marco}, with sign fixed by \eqref{soleta} in the gauge in which it is a pure boost. 
If there is a second critical point as in \eqref{twosols}, the corresponding value of \eqref{normratio2} is obtained simply switching $\eta\mapsto-\eta$, since the additional $\psi$ rotation drops out.
We note also that \eqref{normratio2} is invariant under rotations of the boundary data, as the one for the flag phases. 
It is manifestly independent of rotations of the tilde boundary data. A rotation of the untilded normals rotates the boost direction $\vec u$ as well, hence $\vec{u}\cdot\vec{n}_a$ is invariant.
The vectorial phase can be computed from \eqref{mixedphase}, giving
\be
\bra{\z_a} h^{\dagger}h \sket{\z_a} =-\sinh\eta  \sqrt{1-(\vec u\cdot\vec n_a)^2} \, e^{i \upsilon_a} 
=\frac{\sqrt{j^2_a - k^2_a}}{(1+i \g_a)k_a} e^{i(2\a_a-\tl{\a}_a-\tl{\b}_a)^\cc},
\ee
where
\be
\upsilon_a:=\arg\left( \bra{\z_a}u\ra \bra{u}\z_a] \right),
\ee
and $\ket{u}$ indicates an SU(2) coherent state \eqref{plat} constructed from the direction of the boost. For example
if the boost is aligned with the $\hat{z}$ axis,  $\upsilon_a=\phi(\vec{n}_a) + \pi$, where $\phi(\vec{n}_a)$ is the azimuthal angle of $\vec{n}_a$.
Plugging in the solution \eqref{Marco} for the rapidity, we find
\begin{equation}
\label{mixedphase1}
(2\a_a-\tl{\a}_a-\tl{\b}_a)^\cc=\left\{ 
\begin{array}{lr}\upsilon_a+\arctan\g_a +\pi& {\rm if\ }\eta>0 \\ \upsilon_a+\arctan\g_a & {\rm if\ }\eta<0 \end{array}
\right.
\end{equation}
Here we assumed $\g_a>0$. If it is negative, the arctan is shifted by  $\pi$. This gives the same solutions but with relation with the sign of $\eta$ reversed.

Using these expressions, the on-shell action is 
\be
\Phi^\cc:=\im(S^\cc) =\sum_{a}\r_{a}\log(\cosh \eta -\vec{u}\cdot\vec{n}_a \sinh \eta ) -k_a(\upsilon_a+\arctan\g_a+\f{1+\sgn\,\eta}2\pi) + j_a(\tl\a_a-\tl\b_a)^\cc,
\ee
with the critical flag phase \eqref{critflag}.
When $\g_a=\g$, we know there are two distinct critical points, and 
\begin{align}
\Delta\Phi&: =\Phi^\cc- \Phi^{\Pcc}\nn\\
&=\sum_{a}\g k_a\log\left(\frac{\cosh \eta -\vec{u}\cdot\vec{n}_a\sinh \eta}{\cosh \eta +\vec{u}\cdot\vec{n}_a\sinh \eta} \right) -k_a \pi + j_a
\arg\left(- \frac{\left(\vec{\tl{F}}_a+i\vec{\tl{n}}_a\times\vec{\tl{F}}_a\right)\cdot R \vec{u}\times\vec{n}_a}{\left(\vec{\tl{F}}_a+i\vec{\tl{n}}_a\times\vec{\tl{F}}_a\right)\cdot R R_{\vec{u}}(2\psi) \vec{u}\times\vec{n}_a} \right).
\end{align}
Here $\eta$, $\vec{u}$ and $R$ are the values determined at the critical point $\cc$.\footnote{In writing the sign of the term $-k_a\pi$ we assumed that this has positive rapidity. Inverting this sign has obviously no effect on the formula.} 
As anticipated earlier, this difference is independent of rotations of the two sets of boundary data, and thus of the critical value $r^\cc$ in \eqref{twosols} or the one below. 

Combining all the pieces we can write the asymptotic expansion of the coherent invariant with half-lowest strands,

\begin{align}
\label{lohalf}
{\cal I}^{(\r_a,k_a)}_{j_a \tl\z_i k_a \z_a} 
& = \f{2^{2n+4}\pi^{3-\f n4}}{\l^{3 + \frac{n}{4}}}  \prod_{a=1}^n\left(\frac{j_a^3 k_a^2}{j_a^2-k_a^2}\right)^{\frac 1 4} 
e^{\f i2 \l (\Phi^\cc +\Phi^\Pcc)} \left( N^\cc e^{\f i2 \l \Delta\Phi } + N^{\Pcc} e^{-\f i2 \l \Delta\Phi }\right) +o(\l^{-3-\f n4}).
\end{align}
Here we have collected a global phase to make the rotational-invariant quantity $\D\Phi$ appear explicitly. 
If future explicit computations of the Hessian reveal that $N^\Pcc=\bar N^\cc$ as in the 4-simplex amplitude with lowest spins \cite{noiLor}, or the SU(2) $6j$ symbol, one will recover the cosine leading order of these familiar cases.
This formula is valid for nonsingular normals for $n=3$ and for non-coplanar normals if $n\geq 4$. Looking at the power law, we see that the basic Clebsch-Gordan coefficients (that is, $n=3$) for $\SL(2,\C)$ in the coherent state basis scale like $\l^{-15/4}$, and the boosted coherent tetrahedra (that is, $n=4$) like $\l^{-4}$. An explicit example of critical data and a numerical test of \eqref{lohalf} is provided in the next Section. 

For critical points with $\g_a\neq\g$, we expect a single critical point, as discussed in the previous Section. This does not change the $\l$ scaling, which is the same as in \eqref{lohalf}.

The analysis can now be easily adapted to the case when one or more of the $j_a$ spins takes the lowest value. This happens for instance if we consider polygons that are boosted along a direction aligned with one of the edges. In this case we expect again two distinct critical points, and the analysis is a mixture of the formulas reported in this and the previous Sections. Concerning the asymptotic formula, it will look like \eqref{lohalf}, the main difference being the different scaling factor, which becomes
\be
\label{plow}
\f{2^{2n+4}\pi^{3-\f {(n-p)}4}}{\l^{3 + \frac{(n-p)}{4}}} \prod_{a=1}^p k_a \prod_{a=p+1}^n\left(\frac{j_a^3 k_a^2}{j_a^2-k_a^2}\right)^{\frac 1 4} 
\ee
Here $p$ is the number of lowest spins, and as always we have assumed a non-degenerate Hessian. We will give an example with $n=4$ and an explicit numerical test confirming the fall-off below in Section~\ref{example2}. For $n=3$ and $p=1$, we obtain the scaling $\l^{-7/2}$, which is confirmed by the numerics reported in Fig.~\eqref{Fig3}.

\subsection{General case}
The difficulty with the general case is that we don't know how to solve for $h$ without having to first solve for the flag phases. 
If we try to repeat the previous procedure, and take real and imaginary parts of \eqref{TBB}, both still depend on $h$ and the flag phases: 
\begin{align}
l_a\cosh\eta\,\vec{n}_a+l_a(1-\cosh\eta)\left(\vec{u}\cdot \vec{n}_a\right)\vec{u}_a+
\sinh\eta\,\vec{u}\times \left(\frac{\rho_a k_a}{l_a}\vec{n}_a+\sqrt{l^{2}_a-k^{2}_a}(\vec{p}_a-\frac{\rho_a}{l_a}\vec{q}_a)\right)&=R^{-1}j_a\vec{\tilde{n}}_a
\end{align}
\begin{align}\nn
\cosh\eta \left(\frac{\rho_a k_a}{l_a}\vec{n}_a+\sqrt{l^{2}_a-k^{2}_a}(\vec{p}_a-\frac{\rho_a}{l_a}\vec{q}_a)\right)
-l_a\sinh\eta\,\vec{u}\times \vec{n}_a+&\\
(1-\cosh\eta)\vec{u}\,\vec{u}\cdot \left(\frac{\rho_a k_a}{l_a}\vec{n}_a+\sqrt{l^{2}_a-k^{2}_a}(\vec{p}_a-\frac{\rho_a}{l_a}\vec{q}_a)\right)&
=\frac{\r_a k_a}{j_a}R^{-1}\vec{\tilde{n}}_a+\sqrt{j_a^{2}-k_a^{2}}R^{-1}(\vec{\tilde{p}}-\f{\r_a}{j_a}\vec{\tilde{q}})
\end{align}
In principle, it is possible to use the two independent real equations given by the imaginary part to solve for the flag phases, then plug this in the real part, and solve for $h$. But in practice the solution for the flags is messy, and this gives a non-linear equation in $h$. 
The geometric discussion done previously explain the origin of this difficulty: we cannot use a reference frame to compute the Lorentz transformation, hence we have to simultaneously determine the two frames and the Lorentz transformation relating them. This gives the non-linear equation that we see.
It is therefore just an algebraic difficulty, which does not prevent the geometric interpretation of the critical configurations along the lines previously discussed.
On the other hand, it prevents us from giving an explicit form of the critical action, and to study the multiplicity of distinct critical points. 
We can therefore only give a generic leading order formula, based on the Stirling approximation \eqref{Stirling}:
\begin{align}
{\cal I}^{(\r_a,k_a)}_{j_a \tl\z_a l_a \z_a} &= \f{2^{2n+4}\pi^{3-\f n2}}{\l^{3 + \frac{n}{2}}}  \prod_{a=1}^n\left(\frac{j_a^3}{j_a^2-k_a^2}\right)^{\frac 1 4} \left(\frac{l_a^3}{l_a^2-k_a^2}\right)^{\frac 1 4} \sum_\cc N^\cc e^{i \l \Phi^\cc}.
\end{align}

For completeness, we report the explicit formulas to evaluate the (imaginary part of the) on-shell action, once $h$ and the flag phases have been determined. The vectorial phase can be computed from  \eqref{mixedphase}, explicitly
\begin{align}
\sqrt{j_a^2-k_a^2} e^{i(\a_a+\b_a-\tl\a_a-\tl\b_a)} &= \f{\sqrt{l_a^2-k_a^2}}{2l_a}\Big((l_a-i\r_a)\bra{\z_a}h^\dagger h\ket{\z_a} + (l_a+i\r_a)\sbra{\z_a}h^\dagger h\sket{\z_a}\Big)\\
&\quad + \f{l_a-k_a}{2l_a} e^{i(\a_a-\b_a)}(l_a-i\r_a)\sbra{\z_a}h^\dagger h\ket{\z_a} + \f{l_a+k_a}{2l_a} e^{-i(\a_a-\b_a)} (l_a+i\r_a)\bra{\z_a}h^\dagger h\sket{\z_a}.\nn
\end{align}
And for the norm ratio we have
\begin{align}\label{normratioApp1}
 \f{\norm{z_a}^2}{\norm{h^{-1}z_a}^2} &=\frac{l_a+k_a}{2l_a} \bra{\z_a} h^\dagger h  \ket{\z_a} 
 +\frac{l_a-k_a}{2l_a}\sbra{\z_a} h^{\dagger}h \sket{z_a}  + \frac{\sqrt{l_a^2-k_a^2}}{l_a} \re\Big(e^{-i(\a_a-\b_a)}  \bra{\z_a} h^{\dagger} h\sket{\z_a}\Big).
\end{align}
Both equations can be further simplified using the polar decomposition for $h$ and $\bra{\z_a}h^{\dagger}h \ket{\z_a} =\cosh\eta-\vec{u}\cdot\vec{n}_a\sinh\eta$, $\sbra{\z_a}h^{\dagger}h \sket{\z_a} =\cosh\eta+\vec{u}\cdot\vec{n}_a\sinh\eta$ and $\bra{\z_a}h^{\dagger}h \sket{\z_a} =-\sinh\eta e^{i\upsilon_a} \sqrt{1-(\vec{u}\cdot\vec{n}_a)^2}$. 

\subsection{Boosted coherent states}

The difficulty in solving the general case comes from the fact that one has to determine the flag phases, in order to identify the frame of $\g$-simplicity. The lowest spin case $j=k$ is the only one for which the boundary data provide the complete classical information about the bivectors and their (canonical) frame of $\g$-simplicity. For $j>k$, the spin alone is not sufficient, and that is where the flag phases come in. 
 The information they provide identifies the classical frames most compatible with the fixed quantum labels, and once this is known, the critical group element can be determined. To make an analogy with other problems in spin foam theory, the situation for non-lowest spins is similar to computing the asymptotics of the $\{15j\}$ symbol using orthonormal instead of coherent intertwiners.

Accordingly, the semiclassical analysis becomes much simpler if we increase the amount of classical information carried by the boundary states. This can be done taking boosted SU(2) coherent states, defined by
\be\label{boostedCS}
\ket{\r,k;b,\z} = D^{(\r,k)}(b)\ket{\r,k;k,\z},
\ee
with $b=b^\dagger \in \SL(2,\C)$  a pure boost. The frame of $\g$-simplicity is now explicitly known as the boost $b$ acting on the canonical time-like direction encoded in the lowest spin states. 
The saddle point analysis  with these boundary data is recasted precisely as in the lowest spin case, with
\eqref{magg} replaced by
\begin{align}\label{magg2}
& T^{(1,0)}\big(hb_a) (k_a+i\r_a)\vec{n}_a 
= T^{(1,0)}\big(\tilde b_a) (k_a+i\r_a)\vec{\tl n}_a.
\end{align}
There are no more flag phases to be determined, and the rapidities now appear as labels of the coherent states that select the classical frames of $\g$-simplicity. 
Since this information is now available through the boundary states, we can immediately determine the critical point. 
The solution for non-coplanar data is simply
\be
h^\cc =\tilde b_a r^{\cc}b_a^{-1},
\ee
with the rotation determined as in Appendix~\ref{AppA}, and the critical values for the spinors $z_a$ are determined as in the lowest case. The conditions for existence of solutions is the strand-independence of the right-hand side. The gradient in $h$ produces a closure condition in terms of the boosted bivectors. These two conditions together restore a similar geometric picture to the one discussed so far, with the critical data corresponding to pairs of polygons or polyhedra that can be Lorentz-transformed into one another, with the Lorentz transformation defined by identifying the area normals of the polyhedra as electric parts of $\g$-simple bivectors. All the relevant information about $\g$-simplicity is provided by the data.
The semiclassical limit of the invariants with boosted coherent states is thus significantly simpler, and can be dealt with using the technique of \cite{BarrettLorAsymp} for lowest SU(2) spin. It is the interest from the spin foam perspective to study the problem at fixed (non-lowest) quantum spin that introduces the technical difficulties we addressed in our work.

As a spin-off, our analysis allows us to immediately derive the peak of the spin distribution of the boosted coherent states \eqref{boostedCS},
\be\label{overlap}
\langle\rho,k;j, \tl\z |\rho,k;\beta,\z\rangle=\langle\rho,k;j,\tl\z|D^{(\r,k)}\big(b)|\rho,k;k,\z\rangle.
\ee
The right-hand side contains only the spinorial integration, and its semiclassical limit can be studied with the analysis explained for the half-lowest case, without the part pertaining to the group element gradient.
Writing the boost as $b:=\exp\{\eta \vec u\cdot \tfrac{\vec \s}{2}\}$ and following the saddle point analysis of the half-minimal case, we obtain from \eqref{Tg1} and  \eqref{Marco} that the labels must satisfy
\begin{equation}
\vec{\tilde{n}}= \frac{k}{j} T_{\g}\left(b\right)\vec{n},
\end{equation}
and
\begin{equation}
j^2=k^2 +\left(k^2+\r^{2}\right)\left(1-\left(\vec{n}\cdot\vec{u}\right)^{2}\right)\sinh^2 \eta.
\end{equation}
These conditions select a unique value of $j$ for which the distribution has a dominant fall-off, determined by the irrep and the classical data $(b,\z)$. And introduce restrictions among the boundary data for the existence of such critical behaviour. 
If $\vec u=\vec n$ for instance, the peak is at the lowest value $j=k$ (while the boosted state is still a linear combination of all possible spins), and we must have $\vec n=\vec{\tl n}$.
If the conditions are satisfied, then the solutions for the critical point equations determine the integration spinors as in \eqref{Re}, with the flag phases determined following the procedure of Section~\ref{SecHL} with the parameter $b$ replacing the variable $h$. 
We conclude that there exist critical configurations for which the probability distribution \eqref{overlap} is not exponentially suppressed in the limit of uniformly large $\rho$, $k$, and $j$, and that in this case it peaks at a single point $|\rho,k;j^{(c)},\tl\z^{(c)}\rangle$, determined by our analysis.

\subsection{Spin foams without time gauge}\label{timegauge}
The EPRL model uses only the lowest SU(2) spins.\footnote{Matrix representations with only half the spins in the lowest value occur only when one uses the decomposition of \cite{Boosting} and the booster functions, see Section~\ref{SecBoosters} below.} This simple choice is motivated by the use of the time gauge, which as we have seen, simplifies the description of $\g$-simple bivectors. However, there are precise reasons to be interested in a quantization procedure that does not rely on the time gauge. 
First, because there is no guarantee a priori that the model is
independent of the gauge chosen for the quantization.
Second, even if the model turns out to be gauge-invariant in the bulk, we know that boundaries can transform gauge into physical degrees of freedom (see e.g. \cite{Carlip:1994gy}), hence the relevance of considering the theory in different gauges, or without gauge-fixing altogether.

One way to consider a spin foam model without time gauge, is to study how the simplicity constraints change if the time-like vector is arbitrary, instead of the canonical one $t^I$. At the classical level, a general formula is provided by \eqref{Boostedgsimp}.
At the quantum level, it depends on the details of the map between quantum states and classical bivectors. The map used by the EPRL model in the time gauge is given by \eqref{Pisimple}, and corresponds to the use of lowest spins. We have shown in Section~\ref{sec:gsimple} how this map can be extended to arbitrary SU(2) spins via \eqref{Piprime}, providing a two-parameter family of frames of $\g$-simplicity for each choice of spin via \eqref{FoD}, of which only one is selected by the saddle point equations.  
Therefore a choice of fixed, non-lowest spins corresponds to imposing the  simplicity constraints in an arbitrary gauge, even though the precise frame of this new gauge is only reconstructed after solving the saddle point equations. In concrete terms, this means that it is possible to work with the arbitrary unitary irreps \eqref{Dh} to describe an EPRL-like model without time gauge.

The initial two-parameter freedom can be eliminated, as pointed out in the previous Subsection, if one makes the SU(2) spins semiclassical, working with the boosted coherent states
\eqref{boostedCS}. The classical labels are extended to include information about the frame of $\g$-simplicity, which is now fully identified by the boundary labels, as $N=\L(b)t$. 
If one defines coherent amplitudes for the EPRL model in this way, the simplicity constraints are then implemented in an arbitrary gauge, namely in a   frame that is arbitrarily chosen and classically controlled on each tetrahedron. One can the apply the argument given in \cite{Rovelli:2010ed}, where it was shown that the amplitudes are consistent with local Lorentz invariance in the bulk, and covariance with respect to the boundary data. This result follows  from the definition of the boosted coherent states as group action on the lowest coherent states.

Ultimately, the question of the correct way to implement the simplicity constraints and the role of gauge can only be settled by gaining control on the full semiclassical limit. Our work shows that it is possible to go beyond the use of lowest spins, and how non-lowest spins relate to changing the time gauge.

\section{Examples}
\label{sec:examples}
In this Section we confirm the validity of the saddle point approximation with numerical tests.
We give two examples of critical data, and compare the numerical evaluations of the associated coherent invariants with the asymptotic formulas. 
The numerics are done using a code developed in \cite{Gozzini-to-appear?}, which builds on and improves the code \cite{Dona:2018nev} used in \cite{noiLor}. Wolfram's Mathematica notebooks to evaluate the critical points and on-shell action are available as supplementary material \cite{repo}.
Both examples have $n=4$, strand-independent $\g$ and non-coplanar normals. The spins in one set are at their lowest values, but not in the second set.
This puts us in the half-lowest case of  Section~\ref{SecHL}, for which we know the explicit solution of the critical point equations. 
The critical data are `electric' polyhedra, namely 3d polyhedra that transform under Lorentz as the electric part of $\g$-simple bivectors in the canonical frame. 
As explained in the previous Section, the difficulty with finding an example in the general spin case is that we don't know a priori the frame of definition of $\g$-simplicity, and it is related in a non-linear way to the choice of spins and normals.
For the half-lowest case, our analysis tells us that given a set of lowest spins with normals that close, all that is needed for a critical point is for the second set to be obtained through one of the two solutions \eqref{hsolhm}, namely a specific four-screw, up to a free rotation.
In describing the examples, we proceed as follows: we pick an initial configuration, and we compute the final configuration using \eqref{hsolhm}. The angle and boost parameters are given by \eqref{psiangle} and \eqref{Marco} respectively, and can be determined if we choose $\g$ and the quantum labels of the final configuration, so to respect their domains of definition. But instead of choosing the final normals and determine the direction of the four-screw and the rotation $\rr^\cc$, we can simply choose $\vec u$ and fix $\rr^{\cc}=\Id$, and determine the final configuration. We then use these as boundary data for a numerical evaluation of \eqref{lohalf}. 
As we will see, the numerical results confirms that the configurations are critical, and the predicted asymptotic fall-off and oscillations.

\subsection{Unaligned boost}\label{SecEx1}
For the first example, we start with an equilateral tetrahedron with areas given by the lowest spins $j_a=k_a=k$. We take its area vectors as electric part of $\g$-simple bivectors in the canonical frame.
We consider a four-screw along the axis bisecting one dihedral angle, which we choose by convenience as $\hat z$ axis. 
Using the formula for the rotation angle \eqref{psiangle}, we obtain the following configuration, 
\begin{equation}
\label{case1}
\arraycolsep=1pt
\vcenter{\hbox{\includegraphics[width=2cm]{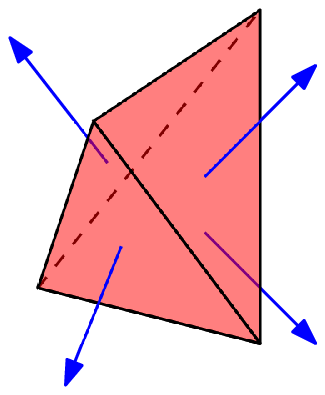} }}
\left\{ \begin{array}{l}
\vec A_1 = k (0                ,   \sqrt{\f{2}{3}}  ,  \f{1}{\sqrt{3}}) \\
\vec A_2 = k (0                , - \sqrt{\f{2}{3}}  ,  \f{1}{\sqrt{3}}) \\
\vec A_3 = k( \sqrt{\f{2}{3}}, 0                  , -\f{1}{\sqrt{3}}) \\
\vec A_4 = k(-\sqrt{\f{2}{3}}, 0                  , -\f{1}{\sqrt{3}})
\end{array} \right.
\qquad \stackrel{\eqref{hsolhm}}{\longrightarrow} \qquad 
\vcenter{\hbox{\includegraphics[width=2cm]{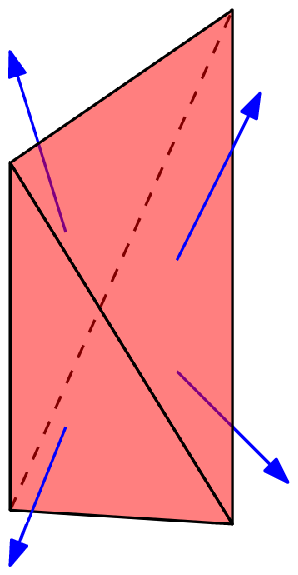} }}
\left\{ \begin{array}{l}
\vec{ \tl A}_1 = j( 0  ,   \sqrt{1-\delta_1^2}, \delta_1 ) \\
\vec{ \tl A}_2 = j( 0    ,  -\sqrt{1-\delta_1^2}, \delta_1 ) \\
\vec{ \tl A}_3 = j(  \sqrt{1-\delta_1^2}, 0, -\delta_1) \\
\vec{ \tl A}_4 = j( -\sqrt{1-\delta_1^2}, 0, -\delta_1)
\end{array} \right.
\end{equation}
where $\d_1:=\frac{k}{\sqrt{3}j}$.  The initial and final tetrahedra are related by a boost with rapidity
\begin{equation}\label{etaex1}
\sinh \eta = \pm \sqrt{\f 3 2}\frac{\sqrt{j^{2}-k^2}}{k\sqrt{1+\gamma^{2}}} \ .
\end{equation}
The sign of the $\psi$ rotation depends on the sign chosen for $\eta$, and both choices give the same final configuration,
whose global frame \eqref{Ncrit} is $N^\cc=(\cosh\eta,0,0,\sinh\eta)$.

The boosted tetrahedron is equi-area but not equilateral, as a consequence of Lorentz contractions. 
Notice however that this transformation is very different from the one that would have incurred if the normals were the spatial part of a four-vector, instead of being the electric part of a $\g$-simple bivector. This is just to say that one has to be careful in attempting to use intuition about Lorentz contractions to find critical boundary data.

Having identified the two critical points $h^\cc$ and $h^\Pcc$ corresponding to the two signs in \eqref{etaex1}, we can compute all remaining quantities.
The flag phases obtained from \eqref{critflag} are
\begin{align}
(\tl \a_1 - \tl \b_1)^\cc &=  f_1 + f_2 - \f{\pi}{2},&  (\tl \a_1 - \tl \b_1)^\Pcc &= f_1 - f_2 +\f{\pi}{2}, \\   
(\tl \a_2 - \tl \b_2)^\cc &=  f_1 + f_2+ \f{\pi}{2}, &  (\tl \a_2 - \tl \b_2)^\Pcc &= f_1 - f_2 -\f{\pi}{2}, \\  
(\tl \a_3 - \tl \b_3)^\cc &=  f_1 - f_2, &  (\tl \a_3 - \tl \b_3)^\Pcc &= f_1 + f_2, \\   
(\tl \a_4 - \tl \b_4)^\cc &=  f_1 - f_2 +\pi, &  (\tl \a_4 - \tl \b_4)^\Pcc &= f_1 + f_2 +\pi ,
 \end{align}
where $f_1= \arctan \left(\frac{k \g}{j}\right)$ and $f_2= \arctan \left(\frac{k \g}{j}\sqrt{\frac{j^2-k^2}{3j^2 + k^2(2\g^2-1)}}\right)$. 
The norm ratios obtained from \eqref{normratio2} are
\begin{align}
\f{\norm {z_1^\cc}^2}{\norm{h^{\cc -1}z_1^\cc}^2} &= \f{\norm {z_2^\cc}^2}{\norm{h^{\cc -1}z_2^\cc}^2} = \log \left(\frac{\sqrt{2 \g^2 k^2+3 j^2 -k^2}-\sqrt{j^2-k^2}}{\sqrt{2} k \sqrt{\g^2+1}}\right), \\
\f{\norm {z_3^\cc}^2}{\norm{h^{\cc -1}z_3^\cc}^2} &= \f{\norm {z_4^\cc}^2}{\norm{h^{\cc -1}z_4^\cc}^2} = \log \left(\frac{\sqrt{2 \g^2 k^2+3 j^2 -k^2}+\sqrt{j^2-k^2}}{\sqrt{2} k \sqrt{\g^2+1}}\right),
 \end{align}
 for the positive sign critical point, and
 \begin{align}
\f{\norm {z_1^\Pcc}^2}{\norm{h^{\Pcc -1}z_1^\Pcc}^2} &= \f{\norm {z_2^\Pcc}^2}{\norm{h^{\Pcc -1}z_2^\Pcc}^2} = \log \left(\frac{\sqrt{2 \g^2 k^2+3 j^2 -k^2}+\sqrt{j^2-k^2}}{\sqrt{2} k \sqrt{\g^2+1}}\right), \\
\f{\norm {z_3^\Pcc}^2}{\norm{h^{\Pcc -1}z_3^\Pcc}^2} &= \f{\norm {z_4^\Pcc}^2}{\norm{h^{\Pcc -1}z_4^\Pcc}^2} = \log \left(\frac{\sqrt{2 \g^2 k^2+3 j^2 -k^2}-\sqrt{j^2-k^2}}{\sqrt{2} k \sqrt{\g^2+1}}\right),
\end{align}
for the other critical point.
The vectorial phases obtained from \eqref{mixedphase1} are
\begin{align}
(2\a_1 - \tl \a_1 - \tl \b_1)^\cc &=  \arctan\g + \f{3}{2}\pi, &  (2\a_1 -\tl \a_1 - \tl \b_1)^\Pcc &=  \arctan\g + \f{\pi}{2}, \\   
(2\a_2 -\tl \a_1 - \tl \b_1)^\cc &=   \arctan\g + \f{\pi}{2},  &  (2\a_2 -\tl \a_2 - \tl \b_2)^\Pcc &=  \arctan\g - \f{\pi}{2}, \\  
(2\a_3 -\tl \a_1 - \tl \b_1)^\cc &=   \arctan\g + \pi,         &  (2\a_3 -\tl \a_3 - \tl \b_3)^\Pcc &=  \arctan\g, \\   
(2\a_4 -\tl \a_1 - \tl \b_1)^\cc &=   \arctan\g + 2\pi,        &  (2\a_4 -\tl \a_4 - \tl \b_4)^\Pcc &=  \arctan\g +\pi .
 \end{align}
 With these values, we find
\begin{equation}
\D \Phi = -4 k \pi.
\end{equation}
Hence, there are no oscillations in the norms of the asymptotic formula \eqref{lohalf}, in spite of the presence of two distinct critical points. This peculiar feature is due to the high symmetry of the configuration chosen in this example, specifically the direction of the boost bisecting a symmetric configuration of normals.

\begin{figure}[h]
\centering
\includegraphics[width=10cm]{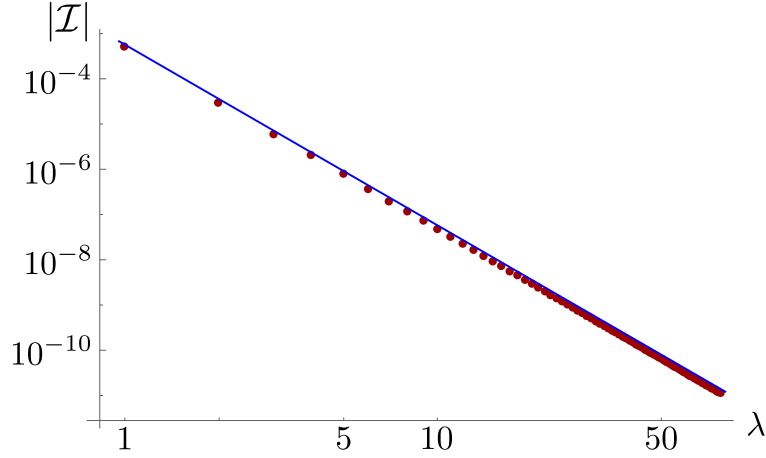}
\caption{\label{figgozzini} \small{\emph{Numerical test of the asymptotic behaviour. The data points are the absolute value of the exact numerical evaluations of the coherent invariant for the configuration of Section~\ref{SecEx1}. The computation was performed with $\g=1.2$, $k=1$ and $j= 2$, using the numerical library \texttt{sl2cfoam-next} \cite{Gozzini-to-appear?}. The straight line is $\propto \l^{-4}$, and it is reported to guide the eye of the reader. The plot confirms the expected power law and absence of oscillations, as explained in the main text due to a numerical coincidence for the symmetry of the boundary data. Picture courtesy of Francesco Gozzini.}} 
} 
\end{figure}

The numerical evaluation of this configuration is presented in Fig.~\ref{figgozzini}. It confirms the validity of the asymptotic formula.

\subsection{Boost aligned with one normal}\label{SecEx2}
\label{example2}
For the second example, we start from the same equilateral tetrahedron, and align the four-screw with one of the normals, which we choose by convenience as $\hat z$ axis. The action of the four-screw preserves this normal and the area of the triangle. We obtain in this way an example in which one of the SU(2) spins is unchanged. 
The transformation gives
\begin{equation}
\label{case2}
\arraycolsep=1pt
\vcenter{\hbox{\includegraphics[width=2cm]{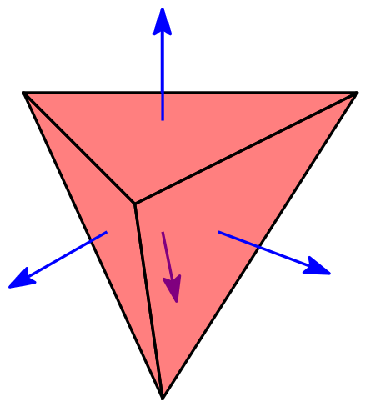} }}
\left\{ \begin{array}{l}
\vec A_1 = k(0      ,  0        , 1        ) \\
\vec A_2 = k(2\f{\sqrt{2}}{3} ,  0              , -\f{1}{3}) \\
\vec A_3 = k(-\f{\sqrt{2}}{3} ,  \sqrt{\f{2}{3}}, -\f{1}{3}) \\
\vec A_4 = k(-\f{\sqrt{2}}{3} , -\sqrt{\f{2}{3}}, -\f{1}{3})
\end{array} \right.
\qquad \stackrel{\eqref{hsolhm}}{\longrightarrow} \qquad
\vcenter{\hbox{\includegraphics[width=2cm]{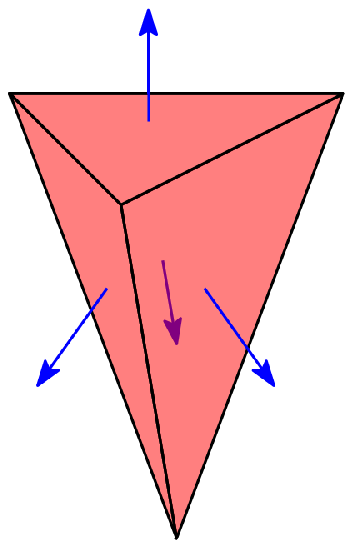} }}
\left\{ \begin{array}{l}
\vec{ \tl A}_1 = k( 0 ,  0    , 1            ) \\
\vec{ \tl A}_2 = j( 2\d_2,  0    , -\frac{k}{3j}) \\
\vec{ \tl A}_3 = j(-\d_2  ,  \sqrt{3} \d_2  , -\frac{k}{3j}) \\
\vec{ \tl A}_4 = j(-\d_2  , -\sqrt{3} \d_2  , -\frac{k}{3j})
\end{array} \right.
\end{equation}
where $\d_2:=\frac{\sqrt{9j^2-k^2}}{6j}$, and the boost that relates the two tetrahedra has rapidity
 \be\label{r1}
\sinh \eta=\pm \frac{3j}{k\sqrt{1+\g^2}} \sqrt{\frac{j^2-k^2}{9j^2-k^2}}.
 \ee
Notice again the role of the bivector map to give a geometric interpretation to the spinors, as opposed to a four-vector map under which the Lorentz contraction of the given tetrahedron would be completely different.

The flag phases of the half lowest strands are obtained from \eqref{critflag} and are
\begin{align}
(\tl \a_2 - \tl \b_2)^\cc &=  f_1 - f_3 + \pi,        &  (\tl \a_2 - \tl \b_2)^\Pcc &= f_1 + f_3 , \\  
(\tl \a_3 - \tl \b_3)^\cc &=  f_1 - f_3 - \f{\pi}{3}, &  (\tl \a_3 - \tl \b_3)^\Pcc &= f_1 + f_3 + \f23 \pi, \\   
(\tl \a_4 - \tl \b_4)^\cc &=  f_1 - f_3 + \f{\pi}{3}, &  (\tl \a_4 - \tl \b_4)^\Pcc &= f_1 + f_3 - \f23 \pi ,
 \end{align}
where $f_1= \arctan \left(\frac{k \g}{j}\right)$ and $f_3= \arctan \left(\frac{k \g}{j}\sqrt{\frac{j^2-k^2}{9j^2 + k^2(8\g^2-1)}}\right)$. 
The norm ratios can be computed from \eqref{normratio2} and gives
\begin{align}
\f{\norm {z_1^\cc}^2}{\norm{h^{\cc -1}z_1^\cc}^2} &= \log \left( \frac{-3\sqrt{j^2-k^2} +  \sqrt{9 j^2 +k^2 (8\g^2-1)}}{2\sqrt{2} k \sqrt{1+\g^2}} \right), \\
\f{\norm {z_2^\cc}^2}{\norm{h^{\cc -1}z_2^\cc}^2} &=\f{\norm {z_3^\cc}^2}{\norm{h^{\cc -1}z_3^\cc}^2}=\f{\norm {z_4^\cc}^2}{\norm{h^{\cc -1}z_4^\cc}^2}= \log \left( \frac{\sqrt{j^2-k^2} +  \sqrt{9 j^2 +k^2 (8\g^2-1)}}{2\sqrt{2} k \sqrt{1+\g^2}} \right),
 \end{align}
and for the other critical point
\begin{align}
\f{\norm {z_1^\Pcc}^2}{\norm{h^{\Pcc -1}z_1^\Pcc}^2} &= \log \left( \frac{3\sqrt{j^2-k^2} +  \sqrt{9 j^2 +k^2 (8\g^2-1)}}{2\sqrt{2} k \sqrt{1+\g^2}} \right), \\
\f{\norm {z_2^\Pcc}^2}{\norm{h^{\Pcc -1}z_2^\Pcc}^2} &=\f{\norm {z_3^\Pcc}^2}{\norm{h^{\Pcc -1}z_3^\Pcc}^2}=\f{\norm {z_4^\Pcc}^2}{\norm{h^{\Pcc -1}z_4^\Pcc}^2}= \log \left( \frac{-\sqrt{j^2-k^2} +  \sqrt{9 j^2 +k^2 (8\g^2-1)}}{2\sqrt{2} k \sqrt{1+\g^2}} \right),
 \end{align}
The vectorial flag of the lowest strands is computed using \eqref{lowestphase},
\begin{align}
(\a_1 - \tl \a_1 )^\cc &= -\frac{1}{2}\arccos\left(\sqrt{\frac{9j^2+k^2(8\g^2-1)}{(9j^2-k^2)(1+\g^2)}}\right), 
& (\a_1 - \tl \a_1 )^\Pcc &=  \frac{1}{2}\arccos\left(\sqrt{\frac{9j^2+k^2(8\g^2-1)}{(9j^2-k^2)(1+\g^2)}}\right),
\end{align}
while the other is computed using \eqref{mixedphase1},
\begin{align}
(2\a_2 -\tl \a_1 - \tl \b_1)^\cc &=   \arctan\g + \pi,         &  (2\a_2 -\tl \a_2 - \tl \b_2)^\Pcc &=  \arctan\g , \\  
(2\a_3 -\tl \a_1 - \tl \b_1)^\cc &=   \arctan\g + \f{5}{3}\pi, &  (2\a_3 -\tl \a_3 - \tl \b_3)^\Pcc &=  \arctan\g + \f{2}{3}\pi, \\   
(2\a_4 -\tl \a_1 - \tl \b_1)^\cc &=   \arctan\g + \f{\pi}{3},  &  (2\a_4 -\tl \a_4 - \tl \b_4)^\Pcc &=  \arctan\g - \f{2}{3}\pi .
 \end{align}
The difference of the phases at the two critical points is
\begin{align}
\D \Phi =& \g k \log \left( \frac{-3\sqrt{j^2-k^2} +  \sqrt{9 j^2 +k^2 (8\g^2-1)}}{3\sqrt{j^2-k^2} +  \sqrt{9 j^2 +k^2 (8\g^2-1)}} \right) + 3 \g k \log \left( \frac{\sqrt{j^2-k^2} +  \sqrt{9 j^2 +k^2 (8\g^2-1)}}{-\sqrt{j^2-k^2} +  \sqrt{9 j^2 +k^2 (8\g^2-1)}} \right) \\ 
& + k \arccos\left(\sqrt{\frac{9j^2+k^2(8\g^2-1)}{(9j^2-k^2)(1+\g^2)}}\right) -3 k \pi + j \pi - 6 j  \arctan \left(\frac{k \g}{j}\sqrt{\frac{j^2-k^2}{9j^2 + k^2(8\g^2-1)}}\right).
\end{align}

\begin{figure}[h]
\centering
\includegraphics[width=10cm]{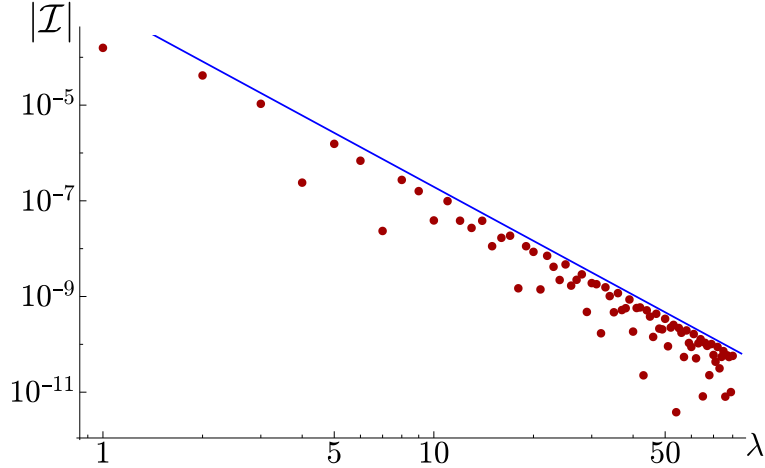}
\caption{\label{figgiorgino} \small{\emph{Numerical test of the asymptotic behaviour. The data points are the absolute value of the exact numerical evaluations of the coherent invariant for the configuration of Section~\ref{SecEx2}. The computation was performed with $\g=1.2$, $k=1$ and $j= 2$, 
using the numerical library \texttt{sl2cfoam} \cite{Dona:2018nev}. The straight line is $\propto \l^{-15/4}$, and it is added to guide the eye of the reader. 
The plot confirms the expected power law and presence of oscillations. Picture courtesy of Giorgio Sarno.}} 
} 
\end{figure}

The asymptotic formula is given by \eqref{plow} with $p=1$ and $n=3$. The corresponding power law is $\l^{-15/4}$, which agrees very well with a numerical calculation of the invariant shown in Fig.~\ref{figgiorgino}.

\section{Coherent tensors and booster functions}
\label{SecBoosters}
Using Cartan's decomposition, the coherent invariant tensors can be written in terms of SU(2) Clebsch-Gordan coefficients and booster functions. 
We have
\be
\label{BandI}
{\cal I}^{(\r,k)}_{j \tl\z l \z} = \sum_{i_n,i_n'} d_{i_n}d_{i_n'}\bar c_{i_n}(\tl\z_a) c_{i'_n}(\z_a) \, B^{(\r,k)}_n(j_a, l_a; i_n,i'_n),
\ee
where:
\be\label{genCG}
c^{j_a}_{i_n}(\z_a) := \sum_{m_a} \vect{j_a}{m_a}^{(i_n)} \bra{j_a,m_a}j_a,\z_a\ra
\ee
is the generating function of Wigner $3jn$ symbols with $n$ legs; 
 $i_n$ is a short-hand notation for the $n-3$ virtual spins that provide a basis of the intertwiner space, with dimension $d_{i_n}$;
and
\be\label{Bn}
 B^{(\r,k)}_n(j_a, l_a; i_n,i'_n)
= \sum_{p_a} \vect{j_a}{p_a}^{(i_n)} \vect{l_a}{p_a}^{(i'_n)} \int_0^\infty d\m(r) \prod_{a=1}^n d^{(\r_a,k_a)}_{j_al_ap_a}(r)
\ee
are the booster functions. See \cite{Boosting} for additional definitions and details.
Conversely,
\be
B^{(\r,k)}_n(j_a, l_a; i_n,i'_n) = \int \prod_{a=1}^n d\m(\z_a)d\m(\tl \z_a) c_{i_n}(\tl\z_a) \bar  c_{i'_n}(\z_a) \, {\cal I}^{(\r,k)}_{j \tl\z l \z}.
\ee

These relations show that the coherent invariant tensors studied in this paper are booster functions in a basis of coherent intertwiner states \cite{LS}. 
It is therefore possible to refer to them as coherent booster functions. They appear in spin foam models if one follows the decomposition of \cite{Boosting} using coherent intertwiners instead of orthonormal intertwiners as edge variables. Such a decomposition can play a useful role in study of the dynamics of spin foam models. It also implies that the Lorentzian asymptotics derived in \cite{BarrettLorAsymp} can be understood as an interference between the Euclidean asymptotics of the SU(2) $15j$ symbol and the asymptotics of the booster functions. This interference can be now studied explicitly thanks to \eqref{lohalf} and the on-shell action we computed. 

We conclude with a comment on the relation between the asymptotics of the coherent boosters studied here, and the asymptotics of the (orthonormal) boosters \eqref{Bn}. The two asymptotics cannot be simply computed from one another, because the generating functions \eqref{genCG} have non-trivial asymptotics themselves.
The  asymptotics of the orthonormal boosters \eqref{Bn} have been estimated analytically only in the case of lowest spins, finding the universal behaviour $\l^{-3/2}$ for any $n\geq 4$ and non-coplanar data \cite{Puchta:2013lza}. This was confirmed numerically in \cite{Boosting}, where the case $n=3$ was also studied. The latter presents two different scalings, 
\begin{align}\label{boosterasympt3}
B^{(\l\r,\l j)}_3(\l j_a, \l j_a) \sim \left\{ \begin{array}{ll} \l^{-3/2} & {\ \rm if \ }\sum_a j_a = 2N+1 \\ \l^{-1} & {\ \rm if \ }\sum_a j_a = 2N \end{array}\right.
\end{align}
The second, dominant scaling is due to the existence of  singular configurations with all normals aligned, which is only possible for even total spin because of the closure condition. The $n=3$ scaling can be easily verified to be consistent with the one obtained for the coherent booster functions.
 Simple numerical tests show that in the 3-valent case
\begin{align}\label{Fscaling}
c^{\l j_a}(\z_a)  \sim  \left\{ \begin{array}{ll} \l^{-3/4} &  {\ \rm if \ } \sum_a j_a = 2N+1 \\ \l^{-1/2} & {\ \rm if \ }\sum_a j_a = 2N \end{array}\right.
\end{align}
Plugging the scalings \eqref{boosterasympt3} and \eqref{Fscaling} in the formula for the coherent tensors \eqref{BandI} we recover the power law of \eqref{LOms}.

It would be interesting to study the relation between the asymptotics of the orthonormal and coherent boosters further. See for instance if it makes it possible to improve on the results of \cite{Puchta:2013lza}: the dependence on the intertwiners of the leading order term in the asymptotics there found is in fact a simple Kronecker delta, which comes drastically short of the actual Gaussian distribution exposed by numerical investigations \cite{Boosting, Dona:2018nev, noiGen}.
Another question related to applications for the EPRL model is to identify the interference pattern in the half-lowest case, which may suggest useful truncations in numerical evaluation of the full amplitude.
Work in these directions will necessarily pass through a detailed analysis of the overlap with the generating functions.
To that end, it is useful to recall that some asymptotics of the generating functions \eqref{genCG} have been explored in \cite{EteraHoloQT,Bonzom:2012bn}, and a possible connection to SO(4) invariants is being studied in \cite{PietroSO4}.

\section{Conclusions}

We have shown how the semiclassical analysis of  \cite{BarrettLorAsymp} can be extended beyond the case of lowest SU(2) spin. The main technical tool is the factorization property of coherent states \eqref{Fcs}, which holds for any spin. Standard $\g$-simple bivectors, satisfying \eqref{gsimp} in the canonical time-like frame, occur only for the lowest SU(2) spin. 
The notion that emerges naturally from the saddle point approximation in the general case is that of boosted $\g$-simple bivectors, satisfying \eqref{Boostedgsimp}. The time-like frame of the boosted $\g$-simplicity depends on the SU(2) spin but requires also additional data. Solving the critical point equations provides the needed additional data, and determines the time-like direction with respect to the reference frame provided by the flags of the boundary spinors. The spinors' flags, or equivalently the choice of global phase of the SU(2) coherent states, acquire thus a more prominent role than they have in the lowest spin case.
In other words, the critical point equations are harder to solve in the general case because the SU(2) coherent states don't contain enough classical information to identify the frame of $\g$-simplicity, as they do in the lowest-spin case. We have also shown how the missing information can be included - and the analysis simplified - if one modifies the asymptotic problem by using boosted SU(2) coherent states, instead of SU(2) coherent states with non-lowest spin.

Our analysis explains how to endow the boundary data within a covariant Minkowskian picture. On each strand, the irrep labels and the SU(2) spin identify a two-parameter family of unit time-like normals $N^I_a$. 
Only if the spin is lowest the degeneracy in the frame of $\g$-simplicity can be eliminated, and the canonical time-like normal singled out.
The spin together with the spinor's null pole define a 3d vector $j_a \vec n_a$ which we identify as the electric part of a bivector which is $\g$-simple in the frame of $N^I_a$. For non-singular configurations, the boosted orientation equations \eqref{magg} fix uniquely these frames in terms of the spinors' flags and the requirement that a unique Lorentz transformation exists between the two sets. The closure conditions \eqref{closure} restricts the 3d vectors to configurations describing polygons or polyhedra.
Stated in spin foam terms, it is the canonical time gauge used in the definition of the $Y$-map that allows the spins of the boundary states of the EPRL model to be interpreted univocally as areas. This has the effect of making the role of the spinors' flags trivial. 
The use of time gauge has been shown not to hinder local Lorentz invariance in the EPRL model \cite{Rovelli:2010ed}, yet constructing a Lorentzian spin foam model that goes beyond the time gauge allows one to test the consistency of the approach
and to study interesting boundary observables. Our results show how the use of non-lowest spins is related to changing the frame of imposition of the simplicity constraints, and we expect them to find applications in this research direction. 
The most immediate application of our results is within the current definition of the EPRL model, where we provide an asymptotics for the (coherent) booster functions, and setting the stage to study the quantum dynamics in terms of interference between the SU(2) vertex amplitudes and the booster functions on the edges, as explained in Section~\ref{SecBoosters}.

Another novelty of the situation with non-lowest SU(2) spin is that the real part of the action at the dominant critical points is not zero, but strictly positive. This quantity however combines with a non-trivial prefactor, and the final leading order asymptotics is purely power-law. As in the lowest spin case, the study of critical points beyond the dominant ones is harder, with more complicated equations to be solved. The general bivector interpretation described in this paper may however be of use also for such analysis.
In this paper we did not explicitly compute the Hessian matrix. This can be done adapting the algorithm for  lowest spins described in \cite{BarrettLorAsymp} and more explicitly in \cite{noiLor} (see also \cite{Han:2020fil}). The numerics performed here are sufficient to deduce that that the determinant at the critical point is non-singular.

Our formula \eqref{Boostedgsimp} provides a compact description of $\g$-simplicity in an arbitrary time-like frame. 
The notion of $\g$-simplicity can be formulated also for space-like or light-like frames (this can be elegantly done using twistors, see \cite{IoWolfgang,IoMiklos}), and this is relevant if one wants to extend the analysis presented here to $\SL(2,\C)$ invariant tensors in terms of other little group  basis rather than Naimark's SU(2) one, see also \cite{Conrady:2010kc,Kaminski:2017eew}.

\subsection*{Acknowledgments}
We thank Giorgio Sarno and Francesco Gozzini for discussions, and for providing us with the data used for the numerical tests presented here. The work of P.D. is supported by the grant 2018-190485 (5881) of the Foundational Questions Institute and the Fetzer Franklin Fund. The work of P.M.D. is supported through grant ID\# 61466 from the John Templeton Foundation, as part of the ``Quantum Information Structure of Spacetime (QISS)'' project (qiss.fr). The opinions expressed in this publication are those of the authors and do not necessarily reflect the views of the John Templeton Foundation.

\begin{appendix}
\setcounter{equation}{0}
\renewcommand{\theequation}{\Alph{section}.\arabic{equation}}

\section{The rotation or pure boost between two electric polygons is unique}
\label{AppA}
Given two unit vectors, there is a two-parameters family of rotations  mapping one into the other, spanned by arbitrary rotations along the directions of the two vectors. Given two sets of $n$ vectors related by a rotation, $R\vec{n}_a = \vec{\tl n}_a$, the rotation is necessarily unique. This holds also for coplanar vectors, provided only they are not all aligned. Uniqueness of the rotation can be proved using  Rodrigues' rotation formula \eqref{rodrotation}
\begin{equation}
\label{rodrig}
R\vec{n}_a =  \cos\th\vec{n}_{a}+\sin\th\vec{v}\times\vec{n}_{a}+(1-\cos\th)\vec{v}(\vec{v}\cdot\vec{n}_{a}) = \vec{\tl n}_a,
\end{equation}
and solving for $\vec v$ and $\th$ in terms of the normals. From this equation and the fact that $R\in SO(3)$ we immediately see that $\norm{\vec{n}_{a}}=\norm{\vec{\tl n}_{a}}=1$. Projecting \eqref{rodrig} along $\vec{v}$ we obtain $\vec{v}\cdot\vec{n}_{a}	=\vec{v}\cdot\vec{\tl n}_{a}$. This has to be true for any $a$, therefore
\begin{equation}
\label{direction}
\vec{v} \propto (\vec{n}_a - \vec{\tl n}_a)\times (\vec{n}_b - \vec{\tl n}_b)
\end{equation}
for any pair $a,b$.
To determine the angle, we take the scalar product of \eqref{rodrig} with $\vec{n}_b$. 
Taking the symmetric and antisymmetric part of this equation in $ab$ we find
\begin{equation}
\label{angle}
\cos\th	= \frac{\vec{n}_{b}\cdot\vec{\tl n}_{a}+\vec{n}_{a}\cdot\vec{\tl n}_{b}-2\vec{v}\cdot\vec{n}_{b}\vec{v}\cdot\vec{n}_{a}}{2\vec{v}\times\vec{n}_{b}\cdot\vec{v}\times\vec{n}_{a}} \ ,
\qquad \sin\th = \frac{\vec{n}_{b}\cdot\vec{\tl n}_{a}-\vec{n}_{a}\cdot\vec{\tl n}_{b}}{2\vec{v}\times\vec{n}_{a}\cdot\vec{n}_{b}} \ .
\end{equation}
If $n>2$ we can compute the direction of the rotation $\vec{v}$ and its angle $\theta$ using any couple of vectors, and their value has to be independent on which couple we choose. This gives us further restrictions on the set of vectors. 
Notice that this proof of unicity holds also for coplanar data.

This construction applies immediately to the normals being the electric parts of $\g$-simple bivectors, since from \eqref{Tgdef} we see that $T_\g(r)=R$. 
For any choice of $a,b$, \eqref{direction} and \eqref{angle} provide an algorithm to compute the critical group element if it is a rotation. Notice that for the solution to exist, the right-hand side of these equations must be independent of the choice of pair $a,b$. These strand-independences are the restrictions on the boundary data necessary for the existence of a rotation relating them.
Consider next a pure boost relating two unit vectors $\vec n_a$ and $\vec {\tl n}_a$ being the electric parts of a $\g$-simple bivector. From \eqref{Tgdef}, we have 
\begin{equation}
\label{rodrigB}
T_{\g_a}(b)\vec{n}_a =  \cosh\eta\vec{n}_{a}+\gamma_a \sinh\eta\vec{u}\times\vec{n}_{a}+(1-\cosh\eta)\vec{u}(\vec{u}\cdot\vec{n}_{a}) = \vec{\tl n}_a.
\end{equation}
If such a boost exists for $a=1,\ldots n$, then it must be unique. This can be proved along the same lines as above, solving explicitly for $b$ in terms of the vectors.
The direction $\vec u$ is determined taking the scalar product of \eqref{rodrigB} with $\vec{u}$ itself. This gives $\vec{u}\cdot\vec{n}_{a}	=\vec{u}\cdot\vec{\tl n}_{a}$. Since this has to hold  for any $a$, we must have
\begin{equation}\label{solu}
\vec{u} \propto (\vec{n}_a - \vec{\tl n}_a)\times (\vec{n}_b - \vec{\tl n}_b)
\end{equation}
for any pair $a,b$.
To compute the rapidity $\eta$,
we take the scalar product of \eqref{rodrigB} with $\g_b\vec{n}_b$. This set of equations can be easily solved to give
\begin{equation}\label{soleta}
\cosh\eta	= \frac{\g_b \vec{n}_{b}\cdot\vec{\tl n}_{a}+\g_a \vec{n}_{a}\cdot\vec{\tl n}_{b}-(\g_a +\g_b)\vec{u}\cdot\vec{n}_{b}\vec{u}\cdot\vec{n}_{a}}{(\g_a +\g_b)\vec{u}\times\vec{n}_{b}\cdot\vec{u}\times\vec{n}_{a}} \ ,
\qquad \sinh\eta = \frac{\vec{n}_{b}\cdot\vec{\tl n}_{a}-\vec{n}_{a}\cdot\vec{\tl n}_{b}}{(\g_a+\g_b)\vec{u}\times\vec{n}_{a}\cdot\vec{n}_{b}}.
\end{equation}
For any choice of $a,b$, \eqref{solu} and \eqref{soleta} provide an algorithm to compute the critical group element if it is a pure boost. Again, the required strand-independences of the right-hand sides are the restrictions on the boundary data necessary for the existence of a pure boost relating them.

The solution so obtained is valid for arbitrary strand-dependent $\g_a$. When they coincide, we can write 
If $\g$ is unique,
\begin{equation}
\cosh\eta	= \frac{\vec{n}_{b}\cdot\vec{\tl n}_{a}+\vec{n}_{a}\cdot\vec{\tl n}_{b}-2\vec{u}\cdot\vec{n}_{b}\vec{u}\cdot\vec{n}_{a}}{2\vec{u}\times\vec{n}_{b}\cdot\vec{u}\times\vec{n}_{a}},
\qquad \sinh\eta = \frac{\vec{n}_{b}\cdot\vec{\tl n}_{a}-\vec{n}_{a}\cdot\vec{\tl n}_{b}}{2\g\vec{u}\times\vec{n}_{a}\cdot\vec{n}_{b}}.
\end{equation}

\end{appendix}
\providecommand{\href}[2]{#2}\begingroup\raggedright\endgroup
\end{document}